\documentclass{statsoc}
\usepackage[a4paper]{geometry}
\usepackage{graphicx}
\usepackage[textwidth=8em,textsize=small]{todonotes}
\usepackage{amsmath}
\usepackage{natbib}
\usepackage{etoolbox}

\makeatletter
\patchcmd{\@makecaption}
  {\parbox}
  {\advance\@tempdima-\fontdimen2} % decrease the width!
  {}{}
\makeatother  

\usepackage[section]{placeins}
\usepackage{amsfonts}            %---Comment if we don't want AMS fonts
\usepackage{amssymb}             %---Comment if we don't want AMS symbols

\usepackage{amsthm}              %---Comment if we don't want fancy theorem environments
\usepackage{amsmath}            %---Comment if we don't want various AMS math goodies
\usepackage{algorithm}           %---Comment if we don't want algorithm support
\usepackage{algorithmic}
\usepackage{stackrel}
\usepackage{float}
\usepackage{placeins}
\usepackage{dblfloatfix}
\usepackage{threeparttable,array,booktabs}
\usepackage{changepage}
\usepackage{url}
\usepackage{natbib}
\usepackage{tikz}
\usetikzlibrary{bayesnet}
\usetikzlibrary{shapes.geometric, arrows}

\title[Static Light Scattering]{Bayesian Analysis of Static Light Scattering Data for Globular Proteins}
\author{Fan Yin}
\address{University of California at Irvine,
USA}
\author{Domarin Khago}
\address{National Cancer Institute, Frederick, USA}
\author[F. Yin, D. Khago, R. Martin, C. Butts]
{Rachel W. Martin, Carter T. Butts\thanks{\emph{Address for correspondence}: Carter T. Butts, Departments of Sociology, Statistics, Computer Science, and EECS
and Institute for Mathematical Behavioral Sciences, University of California at Irvine, SSPA 2145, UCI, Irvine, CA 92697-5100, USA. \newline 
Email:\texttt{buttsc@uci.edu}}}
\address{University of California at Irvine, USA}

\begin{document}
\begin{abstract}
Static light scattering is a popular physical chemistry technique that enables calculation of physical attributes such as the radius of gyration and the second virial coefficient for a macromolecule (e.g., a polymer or a protein) in solution. The second virial coefficient is a physical quantity that characterizes the magnitude and sign of pairwise interactions between particles, and hence is related to aggregation propensity, a property of considerable scientific and practical interest. Estimating the second virial coefficient from experimental data is challenging due both to the degree of precision required and the complexity of the error structure involved. In contrast to conventional approaches based on heuristic OLS estimates, Bayesian inference for the second virial coefficient allows explicit modeling of error processes, incorporation of prior information, and the ability to directly test competing physical models.  Here, we introduce a fully Bayesian model for static light scattering experiments on small-particle systems, with joint inference for concentration, index of refraction, oligomer size, and the second virial coefficient.  We apply our proposed model to study the aggregation behavior of hen egg-white lysozyme and human $\gamma$S-crystallin using in-house experimental data. Based on these observations, we also perform a simulation study on the primary drivers of uncertainty in this family of experiments, showing in particular the potential for improved monitoring and control of concentration to aid inference.  \newline
\end{abstract}

%---Definitions for Defs, Theorems, etc.
\theoremstyle{plain}                        %---Comment out this line if not using amsthm
\newtheorem{axiom}{Axiom}
\newtheorem{lemma}{Lemma}
\newtheorem{theorem}{Theorem}
\newtheorem{corollary}{Corollary}

\theoremstyle{definition}                 %---Comment out this line if not using amsthm
\newtheorem{definition}{Definition}
\newtheorem{hypothesis}{Hypothesis}
\newtheorem{conjecture}{Conjecture}
\newtheorem{example}{Example}

\theoremstyle{remark}                    %---Comment out this line if not using amsthm
\newtheorem{remark}{Remark}

\section{Introduction}
\label{sec:introduction}
For proteins in aqueous solution, measuring association states and propensities towards/away from aggregation is essential for understanding the formation and evolution of both native quaternary structure and deleterious aggregation, due to the fundamental roles of these properties in protein association \citep{bonnete1999second, Bonnete:2002aa, Bolisetty:2011aa, Khatun:2018aa}. Unfortunately, this is difficult, particularly in the highly relevant case of systems at low concentration at or near physiological pH.  Current state-of-the-art approaches (e.g. small-angle X-ray scattering \citep{Hura:2009aa} and neutron scattering \citep{Minezaki:1996aa, Renard:1996aa}) require access to a beamline, which is typically located at a national laboratory or other remote facility. Sending samples to a beamline is expensive and must be scheduled far in advance, which limits the number of sample preparation conditions that can realistically be tested.  A venerable but useful alternative is \emph{static light scattering,} which can allow one to infer such critical quantities as aggregate size (and, in some cases, form factor) and local tendency towards or away from aggregation (as measured by osmotic pressure virial coefficients) \citep{Neal:1998aa, Haas:1999aa}. Unlike X-ray or neutron scattering, light scattering experiments can be performed with commercially available instruments within a typical lab setting \citep{wyatt1993light, Girard:2004aa, Asthagiri:2005aa}, allowing for both greatly reduced cost and greatly enhanced flexibility.

A major barrier to the more widespread use of static light scattering for protein association assays is the lack of a modern, principled approach to data analysis. In the context of soluble proteins (and small oligomers or aggregates thereof), successful inference depends on both error reduction (via a combination of careful experimental procedure and systematic data cleaning) and leveraging of all available information.  Standard approaches within the field, by contrast, are ad hoc and largely depend on graphical techniques developed in the 1940s-1950s \citep{zimm1948scattering}.  These methods provide no principled estimates of uncertainty, and are unable to fully leverage the information content of the available data (e.g., exploiting the consistency between multiple different types of measurements involving the same quantities).  A more modern approach to data analysis could greatly expand the reach of this approach, making it a viable alternative to small-angle X-ray and related measurements for protein research.

In this paper, we address this gap by introducing a systematic approach to the processing and analysis of data from static light scattering experiments on proteins and protein oligomers.  This approach can be generalized to larger protein aggregates and/or other polymers, although we focus on the case of small to medium-sized globular proteins.

Our approach consists of two general elements.  First, we employ a robust data analytic scheme to find and remove experimental artifacts from the data, and to prepare the data for subsequent inferential analysis.  This scheme is intended to be largely automated, with minimal supervision from the analyst required to verify that the data has been properly processed.  Having processed the raw observations, we then employ a hierarchical Bayesian model to correct for known sources of error and infer quantities of scientific interest.  At the core of this model is a joint treatment of light scattering and refractive index data (the latter being required for analysis of light scattering experiments) in a way that allows all available information to be leveraged for inference.

The structure of this paper is organized as follows. The rest of Section \ref{sec:introduction} offers a brief overview on the theory of light scattering and conventional approaches to data analysis. Section \ref{sec:data_cleaning} introduces an automatic procedure for pre-processing the raw data coming directly from the instruments. Section \ref{sec:Bayesian_model} describes the proposed Bayesian model in detail, which is applied to study the aggregates of two soluble, globular proteins: lysozyme and human $\gamma$S-crystallin, in Section \ref{sec:lysozyme_application} and \ref{sec:gammaS_application}, respectively. In Section \ref{sec:simulations}, we conduct simulation experiments to understand the impact of sample size and adjusting for measurement error in concentration measurements on inferential accuracy, the results of which provide critical insights for future experiments. We close with a discussion in Section \ref{sec:discussion} and conclusions in Section \ref{sec:conclusion}.

\subsection{Small Particle Scattering: Theoretical and Experimental Background}
Static Light Scattering (SLS) provides information regarding (variously) mass, radius of gyration, or interaction propensity among particles in solution, by exploiting the way in which these properties affect the scattering of incident light \citep{Attri:2005aa, Minton:2007aa, Fernandez:2009aa}.  Specifically, if a sample is illuminated by a beam of light at a fixed angle and wavelength, the intensity of light scattered at some angle $\theta$ relative to the angle of incidence is a function of the properties of the scatterer, allowing the latter to be inferred from the former.  This intensity is usually referred to in terms of the \emph{Rayleigh ratio}, $R_{c,\theta}$, an observable function of the intensity of the light detected at angle $\theta$ relative to the intensity of the incident beam \citep{Moreels:1987aa}. In practice, the Rayleigh ratio depends upon the concentration of the solute ($c$), among other quantities; while its exact behavior is complex, for the specific case of small particles in dilute solution it can be approximated as \citep{zimm1948scattering, wyatt1993light}
\begin{align}
\label{eq:Rayleigh_ratio}
R_{c,\theta} & \approx K^{*} M_{w} P(r_{g}, \theta) c [1 - 2A_{2}M_{w}P(r_{g}, \theta)c] + O(c^{2}) \\ \nonumber
& = K^{*} M_{w} P(r_{g},\theta) c - 2K^{*}A_{2}M_{w}^{2}P(r_{g},\theta)^{2}c^2 + O(c^{2}),
\end{align}
\noindent or alternatively in its reciprocal form as \citep{hiemenz2007polymer}
\begin{align}
\label{eq:Rayleigh_ratio_reciprocal}
\frac{K^{*} c}{R_{c,\theta}} \approx  \frac{1}{M_{w} P(r_{g}, \theta)} + 2 A_{2} c + O(c^{2}) 
\end{align}
\noindent where $P(r_{g}, \theta)$ is an angular dependence factor, $c$ is the concentration of the solute, $K^{*}$ is a material constant, $M_{w}$ is the \emph{weight} averaged molecular weight of the scattering particle, and $A_2$ is the second virial coefficient, a key physical constant of substantial scientific interest, governing the strength of pairwise interactions among particles. An intuitive explanation of \eqref{eq:Rayleigh_ratio} is that the scattering intensity can be approximated to first order by the mass and concentration of particles from which the beam can scatter (linear term), with a second-order effect arising from the the pairwise interactions among particles (quadratic term): particles that tend to cluster ($A_2<0$) act ``larger,'' on average, generating a stronger signal, while particles that avoid each other ($A_2>0$) produce fewer clusters and lower scattering intensity.  Higher-order virial coefficients (e.g., $A_3$) govern the contributions from higher-order interactions between molecules; in dilute solution, such effects are small and exceedingly difficult to measure, and as such their contribution is generally discarded.

In addition to the affect of concentration and particle interaction, the Rayleigh ratio depends upon two other factors. The material constant $K^{*}$,
$$ K^{*} \equiv \frac{4 \pi^2 n_{0}^2 (dn/dc)^2}{N_{A} \lambda^4}$$
\noindent depends on the intrinsic properties of the materials used for the experiments, and of the light source: the wavelength of the incident light $\lambda$, the refractive index of the solvent $n_{0}$, the refractive index increment, i.e. $dn/dc$ of the solute/solvent pair \footnote{ Note $n_c$, $n_0$ and $dn/dc$ satisfy the following physical constraint, $\frac{n_{c} - n_{0}}{c} = dn/dc$, or equivalently, $n_c = n_0 + c \times dn/dc $}, the mathematical constant $\pi$, and a physical/chemical constant $N_{A} = 6.022 \times 10^{23}$ / mol (i.e., Avogadro's number). In general, the intensity of the scattered light also depends on an angular dependence factor $P(r_{g}, \theta)$, 
$$ P(r_{g}, \theta)^{-1} \approx 1 + \frac{16 \pi^{2}}{3 \lambda^2} \langle r_{g}^{2} \rangle_{w} \sin^{2}(\frac{\theta}{2}),  $$
\noindent where $\langle r_{g}^{2} \rangle_{w}$ is weight average squared radius of gyration. For large scatterers (e.g., polymers) with radii comparable to the wavelength of the light source, $P(r_{g}, \theta)$ can vary appreciably.  In the case of small particles, however, where $\langle r_{g}^{2} \rangle_{w} \ll \lambda_2$, $ P(r_{g}, \theta)^{-1} \approx 1$ and angular effects can be ignored.

In this paper, we are specifically interested in the use of SLS to study aqueous solutions of non-aggregating globular proteins at low concentration, under illumination by visible light ($\lambda = 657 \times 10^{-7}$ cm).  In this regime, the second-order approximation of Equation~\eqref{eq:Rayleigh_ratio} holds, and we may focus exclusively on pairwise interactions among particles.  Moreover, as these particle sizes are on the order of $10^{-9}$m (i.e., $10^{-7}$cm), angular dependence on scattering is below the detection limit of typical instruments, and we hence take $P(r_{g}, \theta)^{-1} = 1$ throughout.  Without loss of generality, we work with the Rayleigh ratio measured at angle $\theta = 90^{\circ}$.  Since we work at a constant measurement angle (and the regime of interest is not angle-dependent), we henceforth simplify notation by dropping reference to $\theta$ and $P(r_{g}, \theta)$ in the remainder of the paper except as noted otherwise.

\subsubsection{Important Sources of Errors}
Because the second virial coefficient (i.e. $A_{2}$) represents a very small deviation in local effective particle density (relative to uniform mixing), it is challenging to
estimate with high precision. Equation \eqref{eq:Rayleigh_ratio} and \eqref{eq:Rayleigh_ratio_reciprocal} shows that estimating $A_2$ requires knowledge of the concentration $c$, refractive index increment $dn/dc$, and Rayleigh ratio $R_{c}$, all of which are prone to measurement errors of different types and magnitudes. To obtain an accurate point estimate and evaluation of the uncertainty of $A_{2}$, accounting for these measurement errors is of substantial importance. Modeling these errors requires a careful consideration of the experimental procedure used to produce the associated measurements; we discuss this in more detail below.

\subsubsection{Unit of Measurements}
The unit of measurement is key to all physical quantities, and the units of the physical quantities involved in this analysis are listed (or can be derived using those listed) in Table \ref{tb:unit_measurement}. Unless otherwise specified, the units of physical quantities remain the same as listed in Table \ref{tb:unit_measurement} for the rest of this paper. 

\begin{table}
\caption{Units of measurement. The physical quantity with unit of measurement $1$ is unitless. \label{tb:unit_measurement}}
\centering
% \begin{threeparttable}
\fbox{
% \resizebox{\textwidth}{!}{
\begin{tabular}{cll}
% \hline
Physical quantity & Unit of measurement & Description \\ 
\hline
$R_{c,\theta}$ & 1/cm & Rayleigh ratio \\
$n_0$ & 1 & Refractive index of solvent \\ 
$n_c$ & 1 & Refractive index of solute/solvent pair \\ 
$c$ & g/mL & Concentration of of the solute\\ 
$dn/dc$ & mL/g & Refractive index increment \\
$\lambda$ & cm & Wavelength of incident light \\
$A_2$ & mL*mol/g$^2$ & Second virial coefficient \\
$N_{A}$ & $6.022 * 10^{23}$/mol & Avogadro’s number \\
$M_{w}$ & g/mol & Weight average molecular weight  \\
\hline
\end{tabular}
}
% }
% \begin{tablenotes}
% \footnotesize
% \item[\dag] Unitless.
% \end{tablenotes}
% \end{threeparttable}
\end{table}
% Table: name, unit

\subsection{Standard Approaches to Data Analysis}
\label{subsec:standard_approaches}
Conventionally, with refractive index increment ($dn/dc$) and weight average molecular weight $M_{w}$ assumed to be known in advance (or assumed to be accurately measured using other means), and the concentration being measured accurately, SLS data has been analyzed based on the ``Zimm plot,'' a two-stage regression proxy method developed by physical chemists based on Equation~\eqref{eq:Rayleigh_ratio_reciprocal}. Despite its popularity, which is primarily due to simplicity and ease of use, the ``Zimm plot'' cannot provide valid uncertainty estimates and can be numerically unstable.  It can also be sensitive to measurement error, particularly with respect to concentrations (which can be difficult to calibrate precisely); more subtly, concentration enters into estimation of both $dn/dc$ and $A_2$, leading to complex correlations among errors.  Some of these limitations can be mitigated by more principled statistical methods, such as the joint bootstrapped regression combining SLS and refractive index measurements introduced by \citet{prytkova2016multi}.  Although this scheme provides a basis for obtaining confidence intervals for $A_2$, and incorporates the interdependence of $dn/dc$ and $A_2$ estimation, it depends on the assumption of monodispersity (i.e., all scattering particles contain approximately the same number of monomers), and does not offer avenues for incorporation of prior information regarding either estimands or measurement error.  Given, on the one hand, the need to leverage as much information as possible to facilitate precise measurements of $A_2$ from limited experimental data, and, on the other, the availability of substantial physical knowledge regarding model parameters, this last is a consequential limitation. 

The need for a combined treatment of $dn/dc$, $A_2$, and particle size distribution is key to the limitations of heuristic strategies (such as the Zimm plot).  In general, the common assumption that the refractive index increment ($dn/dc$) and weight average molecular weight $M_{w}$ will be known \emph{ex ante} with high precision is not realistic. The refractive index increment, $dn/dc$, is often approximated by a pre-specified constant (a ``standard'' value based on a reference protein such as bovine serum albumin, or some other conventional ``average''), or in more sophisticated cases a theoretical calculation based on refractivity of individual amino acids in solution \citep{mcmeekin1964refractive}; both have been shown to have poor accuracy, especially for proteins such as the crystallins of the eye lens that are selected for high refractive index \citep{khago2018protein}.  In practice, accurate assessment of $dn/dc$ hence requires that it be independently measured with a separate instrument, leading to its own source of measurement error.  The situation for $M_w$ is similar: while some proteins can be safely assumed to be monomeric in dilute solution, SLS is frequently used specifically to investigate proteins that are prone to aggregation and/or the formation of complex oligmeric states.  In such cases, the size of the scattering particles is generally unknown, and even monodispersity may be difficult to guarantee (i.e., one may have a mix of oligomeric states).  While (as we show) $M_w$ can itself be estimated from SLS data, errors in this estimate are obviously intertwined with errors in the estimation of $A_2$ (which depends on it).  As noted above, estimation for all of these quantities depends upon solute concentration, which itself is imperfectly known.  Concentration can be estimated from refractive index data, but this depends on knowledge of $dn/dc$ which, as noted, is itself uncertain.  We thus face a situation in which we have several linked unknowns, which must be resolved by leveraging multiple types of measurements simultaneously.  Procedures that ignore such uncertainties by simply fitting to nominal values of concentration, $dn/dc$, or $M_w$ may lead to seriously biased estimates and misleading uncertainty estimates \citep{carroll1985comparison,gleser1987limiting}. 

% Moreover, measurement errors in explanatory variables may cause a loss of power for 
To incorporate a priori scientific knowledge into the analysis, as well as to effectively account for interacting measurement errors in a principled and unified way, we here propose a fully Bayesian model for analyzing the SLS data based on the idea in \emph{errors-in-variable} (EIV) modeling framework \citep{fuller1987measurement, carroll2006measurement}. This model is specified in details in Section \ref{sec:Bayesian_model}.

\section{Data Cleaning and Preparation} 
\label{sec:data_cleaning}
Before proceeding to a discussion of our model, we first describe procedures involved in cleaning raw SLS data and preparing it for use.  These procedures were employed for the data cases examined here, but are also suggested as a practical workflow for handling data of this kind more generally.  Although our emphasis is on data processing, some comments are also made regarding experimental procedure; it is helpful for the analyst to be aware of these issues both to avoid misinterpretation of experimental artifacts (e.g., spurious transient signals due to air bubbles) and to provide guidance to experimental collaborators regarding elements of procedure that may complicate subsequent data analysis.

\begin{figure}[htp]
    \centering
    \includegraphics[width=0.7\textwidth]{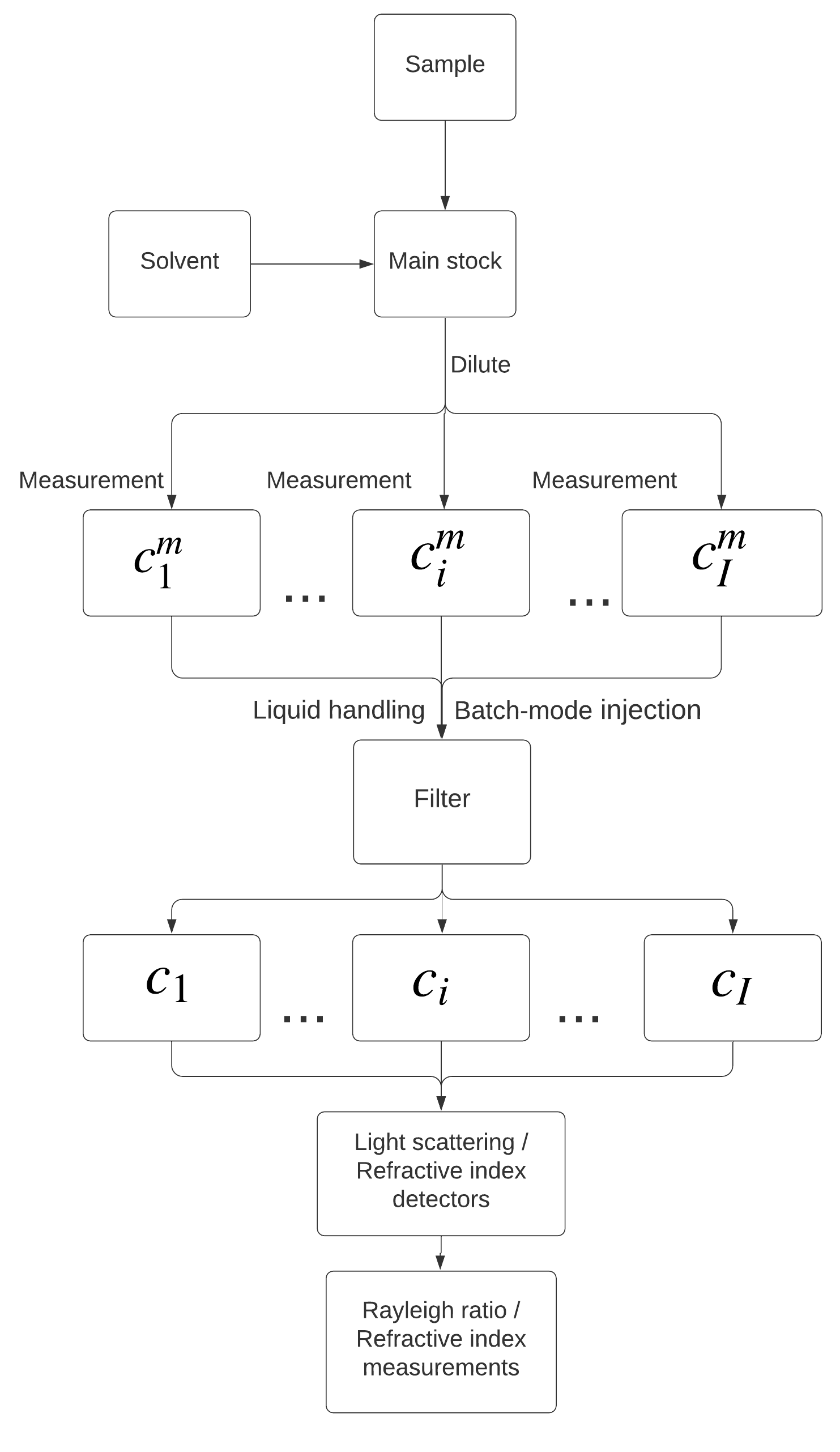}
    \caption{Flowchart showing SLS experimental procedure; labeled quantities are employed in subsequent analyses.}
    \label{fig:SLS_flowchart}
\end{figure}

To obtain the data used in this paper, we performed static light scattering experiments on aqueous solutions of hen egg white lysozyme and human $\gamma$S-crystallin, using a combination of a Dawn HELEOS multi-angle light scattering (LS) detector and an Optilab rEX refractive index (RI) detector from Wyatt Technology (Santa Barbara, CA).  
Figure \ref{fig:SLS_flowchart} shows a flowchart for an SLS experiment. Using batch-mode injections, prepared samples
were injected through a flow cell using a syringe pump starting with buffer only, followed by lysozyme samples (lowest to highest concentration), and finally ending with buffer only for baseline correction. Filters ($0.1 \mu\text{m}$) were used with an aim to enhance monodispersity as samples are injected through the flow cell. Batch-mode collection does introduce potential artifacts causing some inconsistencies in the scattering intensities. These artifacts are results of air bubbles or back pressure from the injection of a new sample during data collection. In this section, we propose an automatic procedure for removing these common and unavoidable artifacts, which prove to greatly facilitate the SLS data preparation process.

\subsection{Removal of Experimental Artifacts} \label{sec_removal}

As noted above, the primary source of experimental artifacts for both light scattering and refractive index measurements are transient changes in signal due to changes in concentration (during the transition from one concentration to another) and/or the presence of air bubbles in the syringe or the flow cell.  In practice, these manifest as large deviations in the input series that are easily noted on inspection of the raw data (see Figure~\ref{fig:LS_pre_cleaning}).  To detect and remove these contaminated observations, we employ a local estimate of the absolute time derivative of the measured signal as a notion of measurement ``stability.''  Specifically, let $y_1,\ldots,y_m$ be a sequence of raw scattering or refractive index measurements at times $t_1,\ldots,t_m$.  We then estimate the instability of measurement $y_i$ by 
\[
\hat{s}_i(y,t) = \left[\frac{|y_{i+1}-y_{i}|}{t_{i+1}-t_i}+\frac{|y_{i}-y_{i-1}|}{t_i-t_{i-1}}\right]/2,
\]
where $|\cdot|$ indicates the absolute value.

To trim transients from the data set, we remove all observations $y_i$ such that
\[
\frac{\hat{s}_i(y,t)-Q_{50}\left(\hat{s}(y,t)\right) }{\mathrm{IQR}\left(\hat{s}(y,t)\right)} > \tau,
\]
where $Q_{50}$ is the 0.5 quantile (i.e., sample median), IQR is the inter-quartile range, and $\tau$ is a user-selected threshold.  We typically select $\tau=1$, which removes all observations whose instability is estimated to be more than one IQR above the median for that time series; in the event that inspection of the trimmed series reveals some ``flyaway'' points remaining, smaller values of $\tau$ may be necessary.

An example of the impact of trimming on raw data series is shown in Figure~\ref{fig:LS_post_cleaning}.  As can be seen, the above criteria have successfully removed most experimental artifacts, leaving a stable interval of points associated with each concentration.  We have found that this approach (with $\tau=1$) typically works quite well, with reductions of $\tau$ to 0.5 or 0 usually sufficing to remove artifacts in difficult cases.  Occasionally, especially dramatic transient fluctuations will yield small clusters of points between concentration levels that are not caught by $\hat{s}$ directly (due the to the fact that the signal fluctuation temporarily ``pauses'').  In these cases, we have found that applying a mild moving average smoothing to $\hat{s}$ by using $\hat{s}'_i(y,t)=(\hat{s}_{i-1}(y,t)+\hat{s}_i(y,t)+\hat{s}_{i+1}(y,t))/3$ in place of $\hat{s}_(y,t)$ readily corrects the problem.

\subsection{Preparation of Raw Measurements} \label{sec_raw_prep}
After removing the experimental artifacts with the procedure described in \ref{sec_removal}, we propose to match the observations with the concentration levels using a clustering algorithm, and then summarize these observations using a robust estimator to assign a single refractive index value and a single Rayleigh ratio value to each concentration level. 

\subsubsection{Clustering of Measurements by Concentration}
Although manual identification of data points associated with each concentration condition is possible, we instead automate the process using a constrained agglomerative clustering algorithm.  %The process proceeds as follows. RWM - I cut this because it is also used below
We initialize the algorithm by taking each set of raw measurements (post-filtering) to be an initial ``cluster;'' we observe that each such measurement is associated with a particular time point, which we exploit in the agglomeration process.  In particular, we consider at each iteration the merger of each current cluster with its immediate temporal neighbors (i.e., the clusters immediately prior and immediately following it in temporal order).  The cost of merging two clusters is determined as follows.  First, the merge penalty for two points is taken to be the median absolute difference between all measurements taken on those points (i.e., if scattering at $k$ angles is observed, the median of the absolute differences over the $k$ angles gives the partial penalty associated with placing these two points within the same cluster.  The merge cost for two clusters is then defined to be the median of the pointwise merge penalties for all point pairs in the respective clusters.  At each stage in the clustering process, the merge cost for each pair of temporally adjacent clusters is computed, and the lowest cost merge is performed.  This process is repeated until the number of clusters is equal to the number of design concentration conditions (plus a concentration zero baseline immediately before and after the non-zero concentrations).

This algorithm is simple to implement and produces solutions with the desired properties of temporal contiguity (a basic feature of the experimental design), repeatability, and optimality with respect to merge cost.  Use of the double median merge cost minimizes the impact of any unstable measurements not removed during the filtering stage on the clustering solution.

\subsubsection{Initial Scattering and Refractive Index Measurements} \label{sec_init_meas}

Given the assignment of (filtered) raw scattering and refractive index measurements to concentration levels, we then obtain initial estimates of $R(\theta,c)$ and $n_c$ by taking the sample median of the raw measurements in each cluster.  Our choice of the sample median is motivated by robustness to remaining unstable measurements; since the number of raw observations per concentration is generally large (e.g., 50-300), efficiency is not a substantial concern.  To obtain a robust initial estimate of the variability of each point estimate, we employ the nonparametric confidence intervals of \citep[][eq. 5.4.10]{gibbons.chakraborti:bk:2011}. (Note that, at this stage in our analysis, we seek an estimate of the variability in the measurement, \emph{not} the underlying parameter.  These are used to set priors for measurement errors within the subsequent stage of analysis.)  

\subsection{Additional Considerations for Practice}

To date, we have found the above approach to work well in handling refractive index and light scattering data taken over a range of concentrations, on multiple instruments.  Typically, little or no manual intervention is required.  However, it is important for the analyst to visually inspect the clustering solution and initial estimates for both scattering and refractive index series to ensure that: (1) unstable observations were properly removed; (2) clustered observations are sensible (and do not e.g. group together observations taken at different concentrations); and (3) initial scattering/refractive index measurements properly represent the central tendencies of measurements at their respective concentration levels.  In our experience, problems with this process invariably result from poor filtering of unstable points.  This can be rectified by modifying the choice of $\tau$ and/or applying moving average smoothing to $\hat{s}$ as described in Section~\ref{sec_removal}.

Although our focus here is on data analysis, some considerations relating to experimental conditions should also be borne in mind.  As we have emphasized above (and will show below), SLS experiments and resulting inference are more sensitive to concentration than many biophysical experiments, and hence precision in sample preparation and liquid handling are a must.  We have also observed that, with multi-angle detectors, it is not uncommon for some detectors to function poorly or to have subtle hardware defects such as loose collimators that induce ``drift'' in measurements over the course of the experiment (observable by inspection as large and often dynamic changes in signal relative to other detectors under simultaneous data acquisition).  Inspection of raw and processed data for bad detectors (and their correction or removal) is hence recommended.  Likewise, it is recommended that the linearity of the RI signal with respect to concentration be verified; we have observed that some refractometers intended for use with SLS experiments will give erroneous readings for proteins with anomalously high $dn/dc$ values, especially at high concentrations, manifesting as large, unphysical, and non-monotone changes in measured RI when the RI exceeds a threshold value.  Removal of RI measurements for these conditions is then essential for accurate analysis. Another important experimental parameter is the temperature, which may produce apparent differences in the refractive index data if uncontrollled \citep{Abbate:1978aa, Tan:2015aa}.

\section{Bayesian Inference for Light Scattering Data}
\label{sec:Bayesian_model}
Given processed data from an SLS experiment, we develop a full Bayesian model for statistical inference for $A_2$, $dn/dc$, and related quantities.  A Bayesian modeling framework is particularly suitable for this problem, as it allows us to efficiently express the complex dependence among the physical quantities in the system, and to incorporate physical information regarding both parameters and the measurement process.  Bayesian model construction also allows us to naturally emulate the logical structure of the experiment itself, with a clear representation of the flow of information from the different measurement processes and priors into the unknown model parameters.  Finally, Bayesian answers for quantities such as posterior uncertainty in $A_2$ values are especially useful given that even high-quality experiments typically estimate $A_2$ with limited precision, and the range of \emph{a posteriori} plausible $A_2$ values is important for tasks such as comparison with simulation studies \citep[e.g.][]{prytkova2016multi}.  Here, we proceed by first describing the structure of the model, followed by prior specification and implementation.  The succeeding sections demonstrate applications to protein data, and provide a simulation experiment probing sensitivity to sample size and data quality.

\subsection{Model Structure}
\begin{figure}[htp]
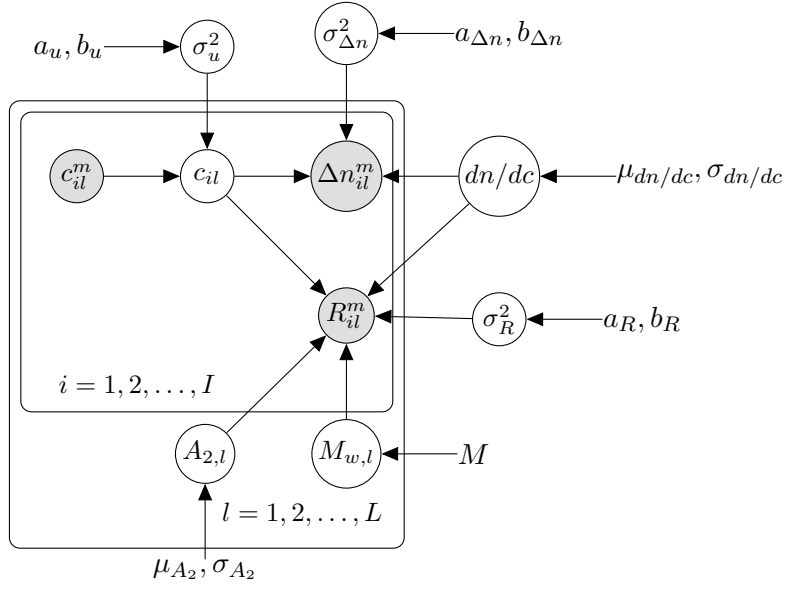

\centering
  \tikz{ %
    \node[obs] (cmil) {$c^m_{il}$} ; %
    \node[latent, right=of cmil] (cil) {$c_{il}$} ; %
    \node[latent, above=of cil] (s2u) {$\sigma_{u}^{2}$} ;
    \node[const, left=of s2u] (s2uprior) {$a_{u}, b_{u}$} ;
    \node[obs, right=of cil] (dn) {$\Delta n_{il}^{m}$} ;
    \node[obs, below=of dn] (R) {$R_{il}^{m}$} ;
    \node[latent, above = of dn] (s2n) {$\sigma^{2}_{\Delta n}$} ;
    \node[const, right = of s2n] (s2nprior) {$a_{\Delta n}, b_{\Delta n}$} ;
    \node[latent, right=of dn] (dn/dc) {$dn/dc$} ;
    \node[const, right = of dn/dc] (dn/dcprior) {$\mu_{dn/dc}, \sigma_{dn/dc}$} ;
    \node[latent, below= of dn/dc] (s2R) {$\sigma_{R}^{2}$} ;
    \node[const, right = of s2R] (s2Rprior) {$a_R, b_R$} ;
    \node[latent, below= of R] (Mw) {$M_{w,l}$} ;
    \node[latent, left = of Mw] (A2) {$A_{2,l}$} ;
    \node[const, below = of A2] (A2prior) {$\mu_{A_{2}}, \sigma_{A_{2}}$} ;
    \node[const, right = of Mw] (M) {$M$} ;
    \plate[inner sep=0.40cm, xshift=-0.12cm, yshift=0.12cm] {Mwplate} {(Mw) (R) (dn) (cil) (cmil)} {$l=1,2,\ldots,L$}; %
    \tikzset{plate caption/.style={caption, node distance=0, inner sep=0pt,
        below left=5pt and 0pt of #1.south,text height=1.2em,text depth=0.3em}} ;
    \plate[inner sep=0.25cm, xshift=-0.12cm, yshift=0.12cm] {conplate} {(cil) (cmil) (dn) (R) } {$i=1,2,\ldots,I$}; %
    \tikzset{plate caption/.style={caption, node distance=0, inner sep=0pt,
          below right=5pt and 0pt of #1.south,text height=1.2em,text depth=0.3em}};
    \edge {cmil,s2u} {cil} ;
    \edge {s2uprior} {s2u} ;
    \edge {cil, s2n, dn/dc} {dn} ;
    \edge {s2nprior} {s2n} ;
    \edge {cil, A2, s2R, Mw, dn/dc} {R} ;
    \edge {M} {Mw};
    \edge {dn/dcprior} {dn/dc} ;
    \edge {A2prior} {A2} ;
    \edge {s2Rprior} {s2R} ;
  }
\caption{\label{fig:f_mod} Structure for the Bayesian SLS model.  Outer plate reflects distinct experimental conditions (e.g., variation in solution conditions), while inner plate reflects measurements at distinct concentrations.  Measured quantities shown as shaded circles, with latent variable as unshaded circles; hyperparameters are shown as uncircled quantities.}
\end{figure}

We assume data in the form of measurements taken under $L$ distinct solution conditions (e.g., ionic concentration, pH, etc.), at $I$ distinct concentrations.  For every condition $l$ and concentration $i$, we observe a concentration measurement $c^m_{il}$, a refractive index measurement (i.e., measured refractive index minus solvent refractive index) $\Delta n_{il}$, and light scattering measurement $R_{il}$.  The plate diagram of figure~\ref{fig:f_mod} shows the structure of the proposed model, which we explain in this section. 

We begin by incorporating known physical constraints. First, we observe that the change in refractive index of the solution (versus solvent), $\Delta n_{il}$, is proportional to the sample concentration, i.e.
\begin{equation}
\label{eq:dRI}
\Delta n_{il} = n_{il} - n_{0l} = c_{il} \times dn/dc
\end{equation}
\noindent where $c_{il}$ is the true concentration (g/mL) corresponding to measured concentration $c^m_{il}$, and $dn/dc$ is the refractive index increment (i.e., $dn/dc$).  We take $dn/dc$ to be constant over the conditions of interest (as is generally the case).  The formula for Rayleigh scattering $R_{il}$ (1/cm) follows that of \eqref{eq:Rayleigh_ratio},
\begin{equation}
R_{il} \approx  K^{*} M_{w,l} c_{il} (1 - 2 A_{2,l} M_{w,l} c_{il} ) \label{eq:R}
\end{equation}
\noindent where $A_{2,l}$ is the second virial coefficient (mol * mL/ g$^{2}$) and $M_{w,l}$ (Da; g/mol) is the mass-weighted scattering unit mass under experimental condition indexed $l$. (For the simplicity of notations, we shall omit the units when specifying the models.)  The material constant $K^{*}$, is given by
$$K^{*} = \frac{4 \pi^2 n_{0}^2 (dn/dc)^2}{N_{A} \lambda^4}.$$ 
\noindent where $n_0$ is the solvent refractive index and $N_A$ is Avogadro's number.  In our experiments, we take the wavelength of the incident light $\lambda = 657 \times 10^{-7}$ cm to be fixed (value determined according to the instrument manual), and we treat $n_{0}$ as fixed because it can be accurately determined by repeated measurements using a high-precision refractometer. Therefore, the only random element of $K^{*}$ is $dn/dc$. 
% where $K$ is a material constant given by the equation \citep{wyatt1993light}
% $$ K^{*} = \frac{4 \pi^2 n_{0}^2 dn/dc^2}{N_{A} \lambda^4}$$
% and $M_{w}$ is the weight average molecular weight (g/mol), $dn/dc$ is the refractive refractive index increment for the sample/solvent pair (mL/g), $N_{A}$ is Avogadro's number (1/mol), $\lambda$ is the incident vertically polarized light in a vacuum (cm), $n_{0}$ is the solvent refractive index, 
% $M_{w,l}$ (Da; $g$/mol)

We model the observed readings of LS and RI detectors, $\Delta R_{il}^{m}$ and $\Delta n_{il}^{m}$, as independent Gaussian and truncated Gaussian random variables centered at the respective theoretical values given by \eqref{eq:R} and \eqref{eq:dRI},
\begin{equation}
\label{eq:R_measurement}
R_{il}^{m} \sim \mathcal{N}(R_{il}, \sigma^{2}_{R})
\end{equation}
\noindent where $\sigma^{2}_{R}$ is the inverse precision of the scattering measurement, and
\begin{equation}
\label{eq:dRI_measurement}
\Delta n_{il}^{m} \sim \mathcal{TN}_{(0,\infty)}(\Delta n_{il}, \sigma^{2}_{\Delta n}) 
\end{equation}
\noindent with $\sigma^{2}_{\Delta n}$ likewise being the inverse precision of the refractive index measurement.  We take these precisions to be constant across measurements.

The measured concentrations $c_{il}^{m}$ are obtained via UV absorption spectroscopy, a high-precision technique.  However, the true concentration $c_{il}$ may still depart from the measured concentration due to the presence of filters and effects from liquid handling (e.g., adhesion of protein to surfaces, effects from transferring the prepared samples from the test tube to the instrument, etc.) that arise after the measurement is made.  To account for these effects, we consider a multiplicative, Berkson-type measurement error model \citep{berkson1950there},
\begin{equation}
\label{eq:c_measurement}
c_{il} = c_{il}^{m} u_{il},
\end{equation}
\noindent where $u_{il}$ is independent of $c_{il}^{m}$ and has a lognormal distribution
\begin{equation}
\label{eq:u_prior}
u_{il} \stackrel{iid}{\sim} \mathcal{LN}(0, \sigma_{u}^{2}),
\end{equation}
\noindent with $\sigma_{u}^{2}$ reflecting the log-variance of concentration discrepancies.

Combining \eqref{eq:R_measurement}, \eqref{eq:dRI_measurement}, \eqref{eq:c_measurement} and \eqref{eq:u_prior} together, we note that our model can be viewed as a multivariate response quadratic regression ($R_{il}^{m}$, $\Delta n_{il}^{m}$) with multiplicative measurement error in the explanatory variable $c_{il}$. It is also worth mentioning that the aggregation status of the protein of interest is unknown (and is often the subject of intense interest) in some experimental settings, and our proposed framework is flexible enough to allow for statistical inference on $M_{w}$;  we discuss this in our sample applications.

\subsection{Prior specification}
\label{subsec:priors}
We assign Gaussian priors to $dn/dc$
\begin{equation}
dn/dc \sim \mathcal{N}(\mu_{dn/dc}, \sigma_{dn/dc}^{2}) \label{eq:dn/dc_prior},
\end{equation} 
\noindent with its location and scale being determined using literature values. To facilitate statistical inference with a parsimonious model, we assume that $dn/dc$ for a specific type of protein is unchanged across different experimental conditions, which is plausible under the conditions covered in our experiments in Table \ref{tb:lysozyme_experiment_conditions}. A hierarchical structure on $dn/dc$ can be adopted when the NaCl concentrations and pH values have larger spans. We then assign inverse-gamma priors to variance parameters, 
\begin{align*}
\sigma_{u}^{2} & \sim IG(a_{u}, b_{u})           \\
\sigma_{R}^{2} & \sim IG(a_{R}, b_{R})         \\
\sigma_{\Delta n}^{2} & \sim IG(a_{\Delta n}, b_{\Delta n})  
\end{align*}
\noindent with the shape and rate parameters are chosen based on precisions reported by the instrument manufacturers and concentrations based on the strength of the prior belief. The prior for the second virial coefficient, $A_2$, is set to be a Gaussian distribution,
$$ A_{2,l} \stackrel{iid}{\sim} \mathcal{N}(\mu_{A_{2}}, \sigma_{A_{2}}^{2}), $$
\noindent where we choose $\mu_{A_{2}} = 0$ and $\sigma_{A_{2}} = 1$ to reflect that $A_2$ can be either positive or negative, but magnitudes greater than $10^{-2}$mL*mol/g$^2$ are extremely unlikely to be physical.

\subsection{Connections with other models}
The proposed model assumes a lognormal-based multiplicative Berkson-type measurement error in the concentrations, which can be viewed as ``explanatory variables'' from a regression modeling standpoint. To the best of the authors' knowledge, this model structure is novel -- it is different from well-established statistical procedures \citep{hwang1986multiplicative, carroll2006measurement} that focus on classical measurement error, and it is also different from the literature on additive Berkson-type measurement error \citep{rudemo1989random, muff2015bayesian} and prior work on bounded multiplicative Berkson-type measurement errors \citep{zhang2012bayesian}.

We note the proposed model is well-posed from a Bayesian perspective in the sense that the posterior distribution of $\sigma_{u}^{2}$ can be estimated as long as the prior for $\sigma_{u}^{2}$ is a legitimate probability distribution \citep{gustafson2005model,gustafson2015bayesian}.  We consider the issue of posterior precision given sample size and data quality in section~\ref{sec:simulations}.
% identifiability issue

\subsection{Implementation}
All computations in this paper were performed in \textbf{R} (version 4.0.1) \citep{R-Team} on a computing server (256GB RAM, with 8 AMD Opteron 6276 processors, operating at 2.3 GHz, with 8 processing cores in each). We used the library R2jags (version 0.6.1) \citep{R2jags} and the JAGS sampler software (version 4.3.0) \citep{plummer2003jags} for conducting MCMC sampling in both case studies and simulation experiments. These are all open-source tools and freely available on the Internet. We note that the proposed model can be implemented using other commonly-used, open-access tools in \textbf{R}, such as \textbf{WinBUGS} \citep{spiegelhalter2003winbugs}, \textbf{Stan} \citep{carpenter2017stan} and \textbf{INLA} \citep{rue2009approximate,rue2017bayesian}.

\section{Application to Aggregation Propensity Assessment in Lysozyme}
\label{sec:lysozyme_application}

In this section, we apply the proposed data cleaning algorithm and the Bayesian model to pre-process and then analyze SLS data collected from experiments on lysozyme, an antimicrobial enzyme produced by animals that forms part of the innate immune system.  Lysozyme can be either aggregation resistant or aggregation prone under particular conditions, and is a common model system for protein aggregation studies \citep{Gripon:1997aa, bonnete1999second, Moon:2000aa}; determining the solution conditions under which $A_2$ switches from positive (repulsive interactions) to negative (attractive interactions) is a point of particular interest.  Here, we examine this question in the context of experiments that vary both pH (altering protonation states, and hence both protein fold and surface charge distribution) and ionic concentration (affecting charge screening, and the stability of salt bridges).   

\subsection{Experimental Conditions and Data Collection}
Lyophilized hen egg white lysozyme was purchased from MP Biomedicals (Solon, OH), and the lysozyme was first weighed and then dissolved in 10 mM sodium phosphate (pH 4.7 and 6.9) containing $0.05\%$ sodium azide and sodium chloride (i.e., NaCl) concentrations at 50, 75, 100, 125, 150, 200, 250, and 300 mM \footnote{a 1 mM (1 millimolar) solution contains 1 millimole per litre (1 mmol/L)} for a target protein concentration of 50mg/mL. This stock solution was then diluted sequentially to produce solutions with nominal lysozyme concentrations of 2.5, 5, 7.5, 10, 12.5, 15, 17.5, 20, 25, 30, 35, 40, 45, and 50 mg/mL, a total of 14 concentration levels. The exact concentrations were measured by UV absorbance spectroscopy using $\epsilon = 2.64 \ \text{mL} \ \text{mM}^{-1} \ \text{cm}^{-1}$ at 280nm (i.e., $2.8 \times 10^{-7}$cm).

% \fan{Hi Domarin, could you please confirm if my interpretation of 1mM is correct? Yes this is correct.}

Table \ref{tb:lysozyme_experiment_conditions} gives a list of experimental conditions. The value in each cell of Table \ref{tb:lysozyme_experiment_conditions} indicates the number of experimental runs under each respective condition. There are two pH levels (4.7, 6.9) and eight NaCl concentration levels (50, 75, 100, 125, 150, 200, 250, and 300 mM) by the original design; however, the data for the experiments under conditions (pH = 4.7, NaCl : 200mM and 250mM) could not be obtained. As a result, the SLS experiments for lysozyme have a total of 14 experimental conditions (varying pH and salt concentration), each of which has one run with 14 lysozyme concentration levels.

\begin{table}
\caption{\label{tb:lysozyme_experiment_conditions} Experimental conditions for lysozyme. Values in each cell indicates the number of experimental runs ($n_0$) under respective condition.}
\centering
\fbox{\resizebox{\textwidth}{!}{\begin{tabular}{l l l l l l l l l}
\hline
NaCl (mM) & 50 & 75 & 100 & 125 & 150 & 200 & 250 & 300 \\
   \hline
pH=4.7 & 1 (1.3272) & 1 (1.3272) & 1 (1.3285) & 1 (1.3279) & 1 (1.3276) & 0 & 0 & 1 (1.3296) \\
pH=6.9 & 1 (1.3291) & 1 (1.3295) & 1 (1.3273) & 1 (1.3289) & 1 (1.3309) & 1 (1.3312) & 1 (1.3331) & 1 (1.3310) \\
\hline
\end{tabular}}}
\end{table}

%   NaCl  pH     n0
% 1   250 6.9 1.3331
% 2   125 6.9 1.3289
% 3   125 4.7 1.3279
% 4    75 4.7 1.3272
% 5    50 4.7 1.3272
% 6   300 6.9 1.3310
% 7   150 6.9 1.3309
% 8    75 6.9 1.3295
% 9   300 4.7 1.3296
% 10  150 4.7 1.3276
% 13  100 4.7 1.3285
% 15  100 6.9 1.3273
% 16   50 6.9 1.3291
% 17  200 6.9 1.3312

\subsection{Data Preparation}
\begin{figure}[htp]
    \centering
    \includegraphics[width=\textwidth]{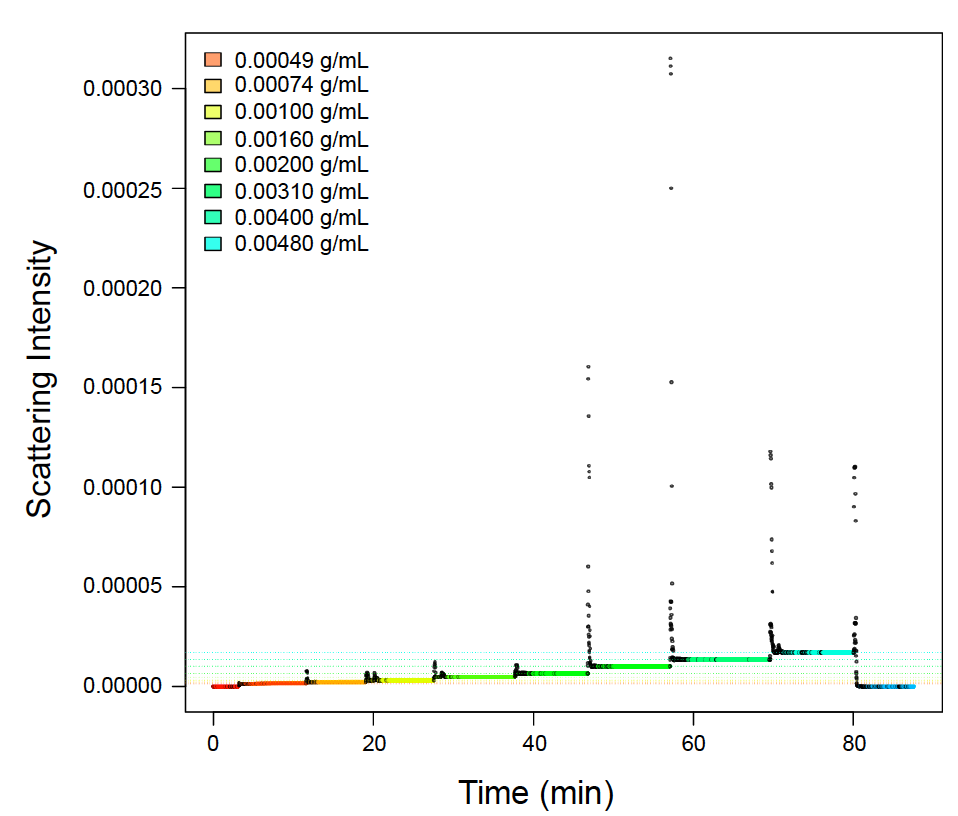}
    \caption{Light scattering data of lysozyme in 10 mM phosphate, 100 mM sodium chloride, $0.05\%$ sodium azide at pH 6.9. Scattering intensity is recorded over the time of the experiment. Each color represents a particular concentration of lysozyme being injected into the MALS instrument, with the first and last being buffer for baseline correction. Black areas indicate artifacts introduced by the sample injection.}
    \label{fig:LS_pre_cleaning}
\end{figure}

\begin{figure}[htp]
    \centering
    \includegraphics[width=\textwidth]{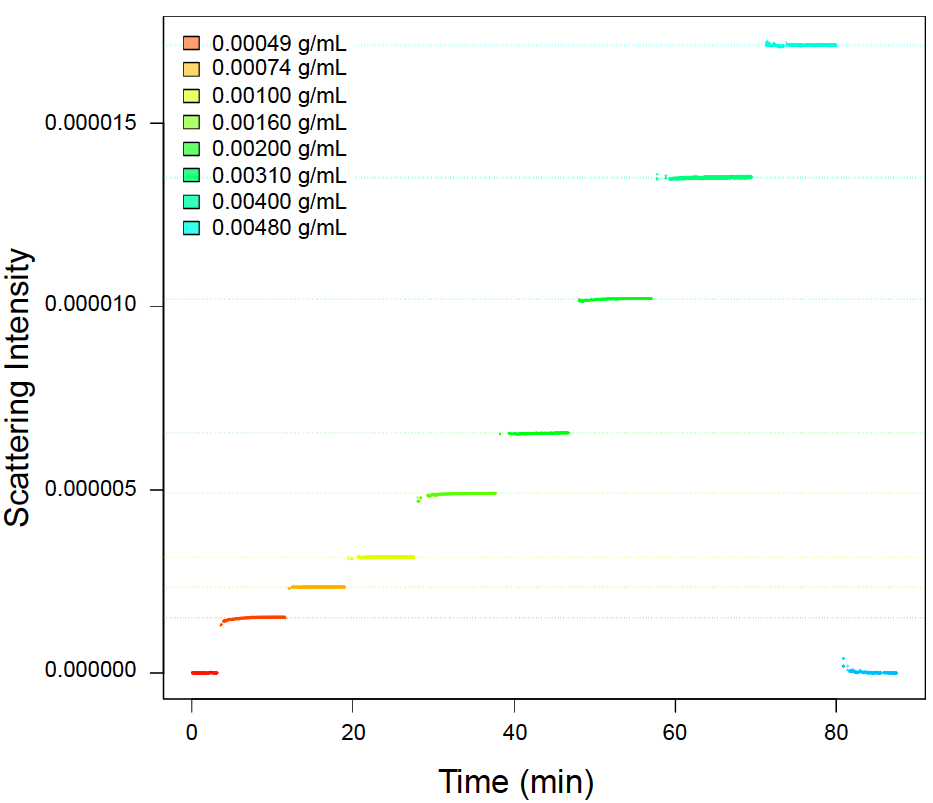}
    \caption{Light scattering data of lysozyme in 10 mM phosphate, 100 mM sodium chloride, $0.05\%$ sodium azide at pH 6.9. Artifacts have been removed.}
    \label{fig:LS_post_cleaning}
\end{figure}

We remove the experimental artifacts, highlighted in black in Figure \ref{fig:LS_pre_cleaning}, using the data pre-processing algorithm developed in Section \ref{sec:data_cleaning}. The pre-processed data are shown in  Figure \ref{fig:LS_post_cleaning} and the median scattering intensity at each concentration was used as the Rayleigh ratio measurement from each detector.

The data cleaning procedure was repeated for all experimental conditions to produce the data for statistical analysis. We also note that the RI detector produced unphysical values for relatively high concentrations, therefore we only included refractive index measurements from  nominal lysozyme concentrations no greater than $20$ mg/mL. All LS measurements are included in the analysis because the LS detector gives physically valid measurements across the entire concentration range after removing the artifacts. Table \ref{tb:lysozyme_LS_RI_measurements} shows which LS and RI measurements are included in the analysis for each experimental run -- within each experimental run, only the first eight concentration levels give valid refractive index measurements, whereas all the concentration levels  provide valid Rayleigh ratio measurements.

\begin{table}
\caption{\label{tb:lysozyme_LS_RI_measurements} LS and RI measurements under different nominal concentration levels within an experiment replicate. ``Y'' indicates the corresponding signal is included in the analysis, otherwise ``N''.}
% \centering
 \fbox{%
  \begin{tabular}{l | l l l l l l l l l l l l l l}
  \hline
nominal concentration level (i=) & 1 & 2 & 3 & 4 & 5 & 6 & 7 & 8 & 9 & 10 & 11 & 12 & 13 & 14 \\
nominal concentration (mg/mL) & 2.5 & 5 & 7.5 & 10 & 12.5 & 15 & 17.5 & 20 & 25 & 30 & 35 & 40 & 45 & 50 \\
   \hline
RI measurements & Y & Y & Y & Y & Y & Y & Y & Y & N & N & N & N & N & N \\
   \hline
LS measurements & Y & Y & Y & Y & Y & Y & Y & Y & Y & Y & Y & Y & Y & Y
   \end{tabular}}
\end{table}

\subsection{Model specification}
Current understanding of the aggregation states of lysozyme in solution under these experimental conditions suggests that oligomers larger than dimers are unlikely to occur under the conditions studied here.  Thus, we propose the following competing models for the aggregation state (expressed via $M_w$):
\begin{itemize}
    \item $\mathcal{M}_{1}$ : $M_{w,l} = M$
    \item $\mathcal{M}_{2}$ : $M_{w,l} = 2M$
    \item $\mathcal{M}_{3}$ : $M_{w,l} = (k_{l}+1) M, \ k_{l} \stackrel{iid}{\sim} Bern(0.5) $
    \item $\mathcal{M}_{4}$ : $M_{w,l} \stackrel{iid}{\sim} \mathcal{U}(M, 2M)$
\end{itemize}
where $M = 14307$ (g/mol) is the molar mass of a lysozyme monomer. The first three models assume monodispersity, that is, the protein molecules are all monomers or dimers within each experimental condition. As it is of substantial interest to explore whether this assumption of monodispersity is supported by experimental data or not, the fourth model relaxes the monodispersity assumption by allowing the weight average molecular weight to take continuous values between the weight of monomers and dimers. A data-driven answer to this scientific question can be facilitated by model selection techniques; specifically, we employ the Deviance Information Criterion (DIC) \citep{spiegelhalter2002bayesian} for this purpose, which can be automatically evaluated by the \textbf{R} function \texttt{bugs} in the package \textbf{R2WinBUGS} \citep{sturtz12r2winbugs}.

To conduct posterior inference, we need to specify hyperparameter values for the prior distribution.  We do so as follows:
\begin{itemize}
    \item $dn/dc \sim \mathcal{TN}_{(0,\infty)}(\mu_{dn/dc} = 0.1970, \sigma_{dn/dc}^{2} = 0.005^2)$, reflecting the prior knowledge that the mean of the refractive index increment of lysozyme is about 0.1970 and the refractive index increment of globular proteins is non-negative and has a range of about $0.03 = 0.005 \times 6$. 
    \item $\sigma_{R}^{2} \sim IG(a_{R} = 1, b_{R} = (10^{-5})^2)$ and $\sigma_{\Delta n}^{2} \sim IG(a_{\Delta n} = 1, b_{\Delta n} = (10^{-4})^2)$, reflecting the prior knowledge that the precision level of LS and RI measurements should have order of magnitude $10^{-5}$ and $10^{-4}$, respectively, while there is also considerable probability that the precision can go beyond or below the nominal level.
    \item $\sigma_{u}^{2} \sim IG(a_{u} = 1, b_{u} = (\log(1+0.05)/1.96)^2)$, reflecting the belief that the true concentration should be between $95\%$ and $105\%$ of the measured value with fairly large probability. 
\end{itemize}

\subsection{Results}
For each candidate model, we run $5$ independent MCMC chains with random starting values and conservative settings ($300000$ total MCMC iterations, burn-in $200000$, storing every $250$-th iteration of the last $100000$ draws as posterior samples). Visual inspection of the trace plots and the Brooks-Gelman-Rubin statistic \citep{gelman1992inference, brooks1998general} shows that the chains mix well.

%  These independent chains are run in parallel, with each chain taking approximately 10.50, 13.56, 12.25, 13.27 minutes for $\mathcal{M}_{1}$, $\mathcal{M}_{2}$, $\mathcal{M}_{3}$ and $\mathcal{M}_{4}$, respectively.

Table \ref{tb:dic_lysozyme} presents the DIC values for the competing models, indicating that $\mathcal{M}_{1}$ and $\mathcal{M}_{3}$ fit the data equally well and are substantially better than other competing models. Further investigations on the posterior samples under $\mathcal{M}_{3}$ show that the $k_{l}, l=1,\ldots,14$ all converge to $0$ (i.e., $M_{w,l} = (k_{l}+1) M = (0+1) M = M$). These results favor the assumption that lysozyme is in the monomeric form under these experimental conditions and hence we select $\mathcal{M}_{1}$ as the model for subsequent inferential analysis on $A_{2}$ due to its simplicity.

Table \ref{tb:A2_post_summary} presents several summary statistics for the posterior samples of $A_{2,l}, l = 1, \ldots,14$. Under each fixed pH, we observe an overall downward trend of $A_2$ values, which is in line with the theory that interactions between monomers become less repulsive as the ionic strength in the solution becomes stronger (i.e., higher NaCl concentrations). Under fixed NaCl concentration, smaller $A_2$ is associated with more neutral environment. Interestingly, this downward pattern is slightly violated when the NaCl concentrations are at 200mM and 250mM (perhaps reflecting a change of conformational state), providing a target for protein structure modeling studies. As shown in the rightmost column of Table \ref{tb:A2_post_summary}, we have high posterior certainty that the pairwise interaction between lysozyme monomers is repulsive (i.e., $P(A_{2}>0|\cdot) \approx 1$) under low NaCl concentrations (pH = 4.7, NaCl: 50, 75, 100, 125 mM; pH = 6.9, NaCl: 50, 75 mM), and fairly high certainty that the pairwise interaction between lysozyme monomers is attractive (i.e., $P(A_{2}>0|\cdot) < 0.01$) under high NaCl concentrations (pH = 4.7, NaCl: 300mM; pH = 6.9, NaCl: 125, 150, 200, 250, 300 mM). These findings confirm the previous experimental observations that high salt conditions promote attractive interaction and hence e.g., crystallization \citep{bonnete1999second}. 

\begin{figure}[htp]
    \centering
    \includegraphics[width=\textwidth]{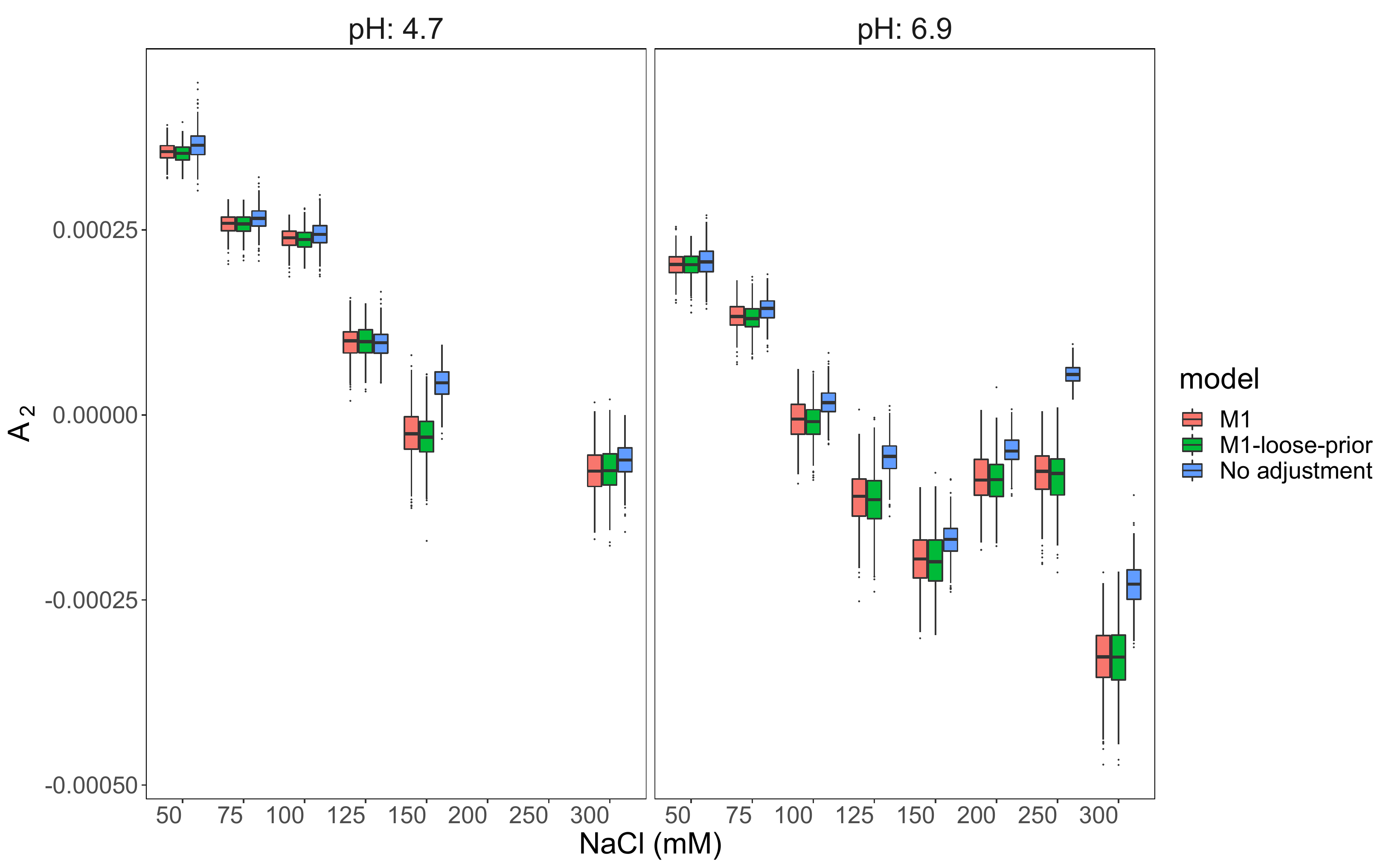}
    \caption{Boxplots of posterior samples: the second virial coefficients ($A_{2}$) estimated from $\mathcal{M}_{1}$ under different priors, and a model without concentration adjustment}
    \label{fig:lysozyme_A2_M1}
\end{figure}

\begin{figure}[htp]
    \centering
    \includegraphics[width=\textwidth]{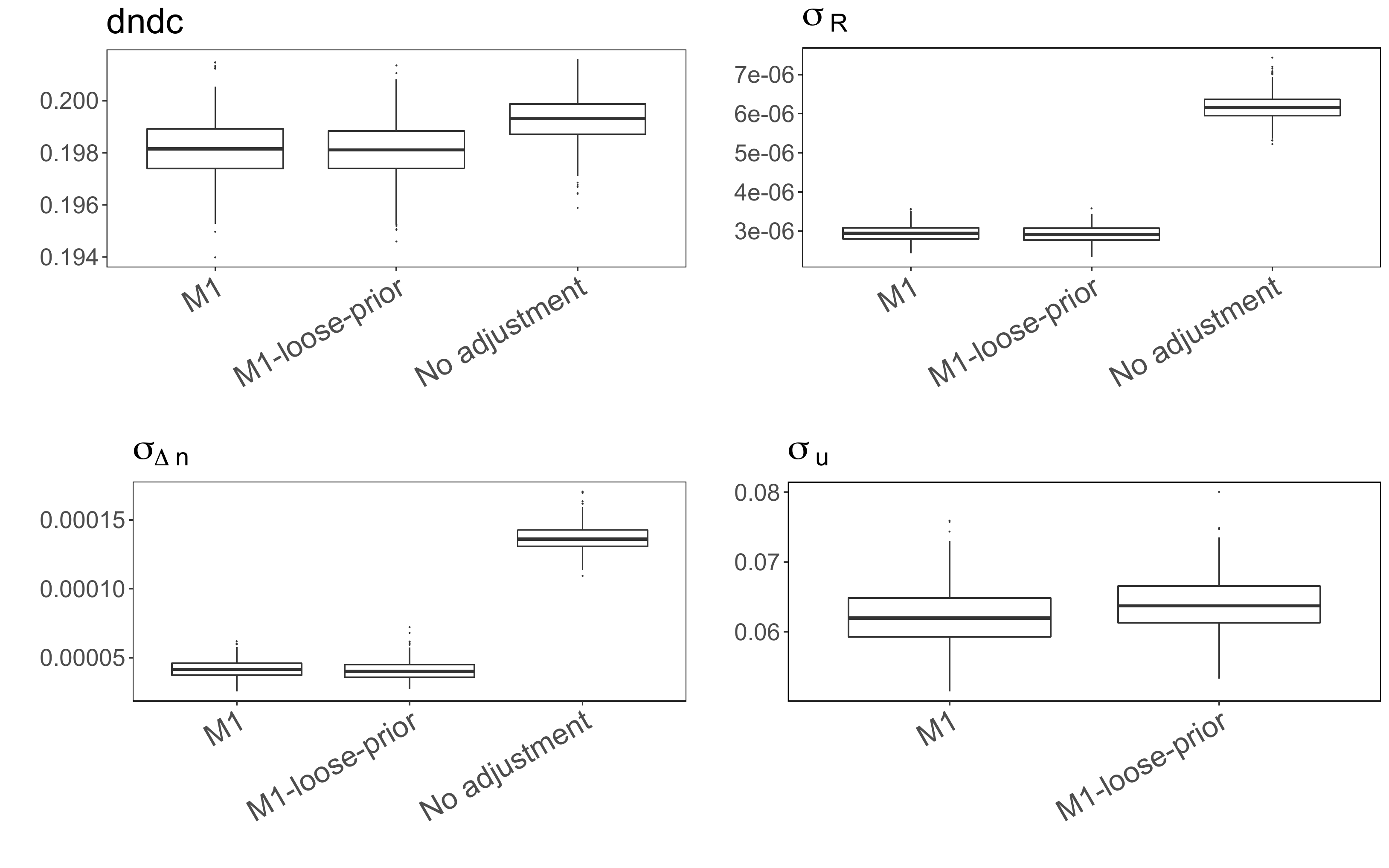}
    \caption{Boxplots of posterior samples: $dn/dc$, $\sigma_{R}$, $\sigma_{\Delta n}$ and $\sigma_{u}$ under $\mathcal{M}_{1}$ and a model without concentration adjustment}
    \label{fig:lysozyme_others_M1}
\end{figure}

We perform a sensitivity analysis with a much looser prior on $\sigma_{u}^{2}$ (i.e., $b_u = (\log(1+0.25)/1.96)^2$) to examine the robustness of our results. Figure \ref{fig:lysozyme_A2_M1} and Figure \ref{fig:lysozyme_others_M1} show the posterior samples of second virial coefficients ($A_2$), $dn/dc$, $\sigma_{R}$, $\sigma_{\Delta n}$ and $\sigma_{u}$. These figures show that our results are in general not sensitive to loose-yet-meaningful priors on $\sigma_{u}^{2}$ (fairly large probability of true concentration falling between $75\%$ and $125\%$ of the measured concentration). As expected, we note that the posterior samples of $\sigma_{u}$ are slightly larger under looser priors. 

As a point of comparison, we also run a model without adjusting for measurement errors in concentrations, in which we treat the measured concentration as the true concentration. Table \ref{tb:A2_post_summary} shows that the model without measurement error adjustment results in very different point and interval estimates of $A_2$, and gives almost opposite qualitative results for the changes in $A_2$ (pH = 4.7, NaCl: 150mM; pH = 6.9, NaCl: 250 mM). As shown in Figure \ref{fig:lysozyme_others_M1}, such a model clearly forces the here unaccounted-for errors in concentration measurements to be propagated into other sources: $\sigma_{\Delta n}$ under the ``no adjustment'' model is estimated to be almost three times that of $\mathcal{M}_{1}$, and $\sigma_{R}$ under the ``no adjustment'' model is estimated to be almost two times that of $\mathcal{M}_{1}$.  These effects emphasize the need to account for concentration errors during analysis.  We further illustrate the importance of adjusting for measurement errors via a simulation study in Section \ref{sec:simulations}.

\begin{table}
\caption{\label{tb:dic_lysozyme} DIC values for candidate models for lysozyme solution. Optimal model with lowest DIC value is highlighted in bold.}
% \centering
 \fbox{%
  \begin{tabular}{c c c c c}
  \hline
   & $\mathcal{M}_{1}$ & $\mathcal{M}_{2}$ & $\mathcal{M}_{3}$ & $\mathcal{M}_{4}$\\ 
%   \hline
    % & $M_{w,l} = M$ & $M_{w,l} = 2*M$ & $M_{w,l} = (k_{l}+1) M, \ k_{l} \stackrel{iid}{\sim} Bern(0.5) $  & $M_{w,l} \stackrel{iid}{\sim} \mathcal{U}(M, 2M)$ \\
   \hline
   DIC & \bf{-5447.9} & -4462.1 & \bf{-5449.3} & -5244.8 \\ 
   \hline
   \end{tabular}}
\end{table}

\begin{table}
\caption{Posterior mean, standard deviation (SD), $2.5\%$ and $97.5\%$ quantile of $A_{2} \times 10^{5}$ under various pH and NaCl strength conditions. The units of $A_{2}$ are mL*mol/$g^2$. The probability of $A_{2}$ being positive is also presented. The numbers in $()$ are the results under the ``no adjustment'' model. \label{tb:A2_post_summary}}
\centering
\fbox{\resizebox{\textwidth}{!}{\begin{tabular}{l|lllllll}
\hline
$l=$ & pH & NaCl (mM) & Mean & SD & $2.5\%$ quantile & $97.5\%$ quantile & $P(A_{2}>0|\cdot)$ \\ 
  \hline
  1 & 4.7 & 50 & 35.521 (36.450) & 1.263(2.090) & 32.859 (32.693) & 37.699 (40.891) & 1.000 (1.000) \\ 
  2 & 4.7 & 75 & 25.803 (26.512) & 1.389 (1.758) & 22.817 (23.089) & 28.237 (29.958) & 1.000 (1.000) \\ 
  3 & 4.7 & 100 & 23.829 (24.431) & 1.417 (1.869) & 20.983 (20.716) & 26.286 (28.052) & 1.000 (1.000) \\ 
  4 & 4.7 & 125 & 9.755 (9.708) & 2.307 (1.986) & 4.894 (5.926) & 14.053 (13.628) & 1.000 (1.000) \\ 
  5 & 4.7 & 150 & -2.681 (4.250) & 3.448 (2.097) & -9.610 (-0.113) & 3.479 (7.788) & $\mathbf{0.217 (0.973))}$ \\ 
  6 & 4.7 & 300 & -7.654 (-6.199) & 3.154 (2.399) & -14.163 (-11.333) & -1.979 (-1.809) & 0.005 (0.000) \\ 
  \hline
  7 & 6.9 & 50 & 20.312 (20.713) & 1.666 (2.138) & 17.281 (16.541) & 23.679 (24.699) & 1.000 (1.000) \\ 
  8 & 6.9 & 75 & 13.308 (14.295) & 1.715 (1.605) & 9.898 (11.341) & 16.435 (17.081) & 1.000 (1.000) \\ 
  9 & 6.9 & 100 & -0.630 (1.688) & 2.840 (1.870) & -6.531 (-2.000) & 4.607 (5.268) & 0.427 (0.835) \\ 
  10 & 6.9 & 125 & -11.225 (-5.739) & 3.588 (2.381) & -18.011 (-10.845) & -4.490 (-1.281) & 0.003 (0.007) \\ 
  11 & 6.9 & 150 & -19.510 (-16.836) & 3.755 (2.564) & -26.754 (-22.297) & -12.304 (-11.917) & 0.000 (0.000) \\ 
  12 & 6.9 & 200 & -8.681 (-4.811) & 3.477 (2.034) & -15.642 (-9.082) & -2.179 (-0.722) & 0.003 (0.007) \\ 
  13 & 6.9 & 250 & -7.888 (5.490) & 3.515 (1.341) & -15.181 (3.083) & -1.318 (8.170) & $\mathbf{0.007 (1.000)}$ \\ 
  14 & 6.9 & 300 & -32.855 (-22.884) & 4.398 (2.989) & -41.955 (-28.556) & -25.435 (-17.275) & 0.000 (0.000) \\ 
  \hline
\end{tabular}}}
\end{table}

\subsubsection{Model assessment}
We conduct posterior predictive checks \citep{gelman1996posterior} to examine whether posterior predictive samples of LS and RI readings can cover the measured values reasonably enough to be scientifically plausible.

Table \ref{tb:lysozyme_LS_post_check} and \ref{tb:lysozyme_RI_post_check} display the posterior predictive $p$-values of Rayleigh ratio and $\Delta n$ readings. Most of the predictive $p$-values are within the range of roughly $0.25$ to $0.75$, suggesting the model provides a good fit to the observed data, especially for $\Delta n$ readings. It is worth noting that the posterior predictive $p$-values for Rayleigh ratio readings are closer to $0.50$ for large concentration levels, and the fit for two experiments (pH = 4.7, NaCl : 150mM; pH = 6.9, NaCl = 250mM) are slightly worse compared to others.  The latter may suggest the presence of unobserved factors associated with experimental procedure affecting these data points.  However, the data are nevertheless quite compatible with what would be expected under the model.

\begin{table}
\caption{Posterior predictive $p$-values for Rayleigh ratio measurements under $\mathcal{M}_{1}$. \label{tb:lysozyme_LS_post_check} }
\centering
\fbox{\resizebox{\textwidth}{!}{
\begin{tabular}{l | llllll | llllllll}
  \hline
NaCl (mM) & 50 & 75 & 100  & 125  & 150  & 300  & 50  & 75 & 100 & 125 & 150 & 200 & 250  & 300  \\ 
    \hline
Concentration level (i=) &  &  & pH = 4.7 &  &  &  &  &  &  & pH = 6.9 &  &  &  & \\
  \hline
  1 & 0.613 & 0.547 & 0.565 & 0.398 & 0.013 & 0.468 & 0.573 & 0.560 & 0.502 & 0.133 & 0.435 & 0.580 & 0.335 & 0.395 \\ 
  2 & 0.605 & 0.610 & 0.630 & 0.435 & 0.007 & 0.515 & 0.490 & 0.588 & 0.490 & 0.122 & 0.443 & 0.525 & 0.185 & 0.463 \\ 
  3 & 0.760 & 0.657 & 0.682 & 0.502 & 0.022 & 0.485 & 0.487 & 0.530 & 0.398 & 0.113 & 0.450 & 0.450 & 0.133 & 0.453 \\ 
  4 & 0.805 & 0.740 & 0.760 & 0.598 & 0.068 & 0.590 & 0.495 & 0.410 & 0.383 & 0.128 & 0.372 & 0.380 & 0.128 & 0.460 \\ 
  5 & 0.890 & 0.780 & 0.738 & 0.677 & 0.140 & 0.527 & 0.570 & 0.335 & 0.335 & 0.128 & 0.432 & 0.285 & 0.077 & 0.435 \\ 
  6 & 0.925 & 0.843 & 0.787 & 0.657 & 0.172 & 0.417 & 0.655 & 0.300 & 0.247 & 0.182 & 0.463 & 0.270 & 0.037 & 0.440 \\ 
  7 & 0.955 & 0.863 & 0.835 & 0.665 & 0.255 & 0.448 & 0.645 & 0.292 & 0.270 & 0.250 & 0.510 & 0.195 & 0.018 & 0.378 \\ 
  8 & 0.985 & 0.895 & 0.790 & 0.757 & 0.020 & 0.465 & 0.760 & 0.280 & 0.180 & 0.338 & 0.580 & 0.177 & 0.013 & 0.340 \\ 
  9 & 0.865 & 0.790 & 0.688 & 0.505 & 0.395 & 0.550 & 0.730 & 0.635 & 0.545 & 0.395 & 0.375 & 0.475 & 0.492 & 0.385 \\ 
  10 & 0.873 & 0.630 & 0.640 & 0.522 & 0.417 & 0.463 & 0.575 & 0.665 & 0.568 & 0.445 & 0.490 & 0.507 & 0.448 & 0.338 \\ 
  11 & 0.682 & 0.613 & 0.540 & 0.487 & 0.510 & 0.422 & 0.420 & 0.593 & 0.545 & 0.440 & 0.485 & 0.487 & 0.415 & 0.448 \\ 
  12 & 0.487 & 0.345 & 0.547 & 0.448 & 0.497 & 0.450 & 0.482 & 0.570 & 0.525 & 0.568 & 0.470 & 0.512 & 0.370 & 0.502 \\ 
  13 & 0.302 & 0.330 & 0.407 & 0.453 & 0.588 & 0.657 & 0.480 & 0.407 & 0.472 & 0.560 & 0.527 & 0.545 & 0.530 & 0.578 \\ 
  14 & 0.115 & 0.330 & 0.328 & 0.502 & 0.693 & 0.435 & 0.420 & 0.425 & 0.455 & 0.555 & 0.555 & 0.485 & 0.787 & 0.662 \\ 
  \hline
\end{tabular}}}
\end{table}

%%%%%%%%%%
\begin{table}
\caption{Posterior predictive p-values for $\Delta n$ measurements under $\mathcal{M}_{1}$. \label{tb:lysozyme_RI_post_check} }
\centering
\fbox{\resizebox{\textwidth}{!}{\begin{tabular}{l | llllll | llllllll}
  \hline
NaCl (mM) & 50 & 75 & 100  & 125  & 150  & 300  & 50  & 75 & 100 & 125 & 150 & 200 & 250  & 300  \\ 
    \hline
Concentration level (i=) &  &  & pH = 4.7 &  &  &  &  &  &  & pH = 6.9 &  &  &  & \\
  \hline
  1 & 0.517 & 0.573 & 0.635 & 0.562 & 0.435 & 0.450 & 0.620 & 0.698 & 0.615 & 0.430 & 0.338 & 0.595 & 0.492 & 0.235 \\ 
  2 & 0.357 & 0.557 & 0.542 & 0.480 & 0.270 & 0.422 & 0.468 & 0.820 & 0.573 & 0.367 & 0.280 & 0.522 & 0.195 & 0.400 \\ 
  3 & 0.378 & 0.480 & 0.475 & 0.545 & 0.385 & 0.448 & 0.547 & 0.728 & 0.570 & 0.410 & 0.328 & 0.578 & 0.407 & 0.468 \\ 
  4 & 0.407 & 0.482 & 0.527 & 0.415 & 0.427 & 0.450 & 0.545 & 0.650 & 0.585 & 0.407 & 0.350 & 0.570 & 0.627 & 0.390 \\ 
  5 & 0.390 & 0.502 & 0.460 & 0.448 & 0.492 & 0.430 & 0.510 & 0.662 & 0.627 & 0.492 & 0.380 & 0.532 & 0.680 & 0.470 \\ 
  6 & 0.450 & 0.455 & 0.450 & 0.482 & 0.515 & 0.480 & 0.522 & 0.610 & 0.603 & 0.472 & 0.422 & 0.570 & 0.720 & 0.475 \\ 
  7 & 0.405 & 0.450 & 0.443 & 0.435 & 0.510 & 0.438 & 0.465 & 0.608 & 0.650 & 0.417 & 0.385 & 0.623 & 0.720 & 0.477 \\ 
  8 & 0.438 & 0.500 & 0.450 & 0.367 & 0.588 & 0.512 & 0.512 & 0.608 & 0.635 & 0.463 & 0.415 & 0.618 & 0.770 & 0.510 \\ 
   \hline
\end{tabular}}}
\end{table}

\begin{figure}[htp]
    \centering
    \includegraphics[width=\textwidth]{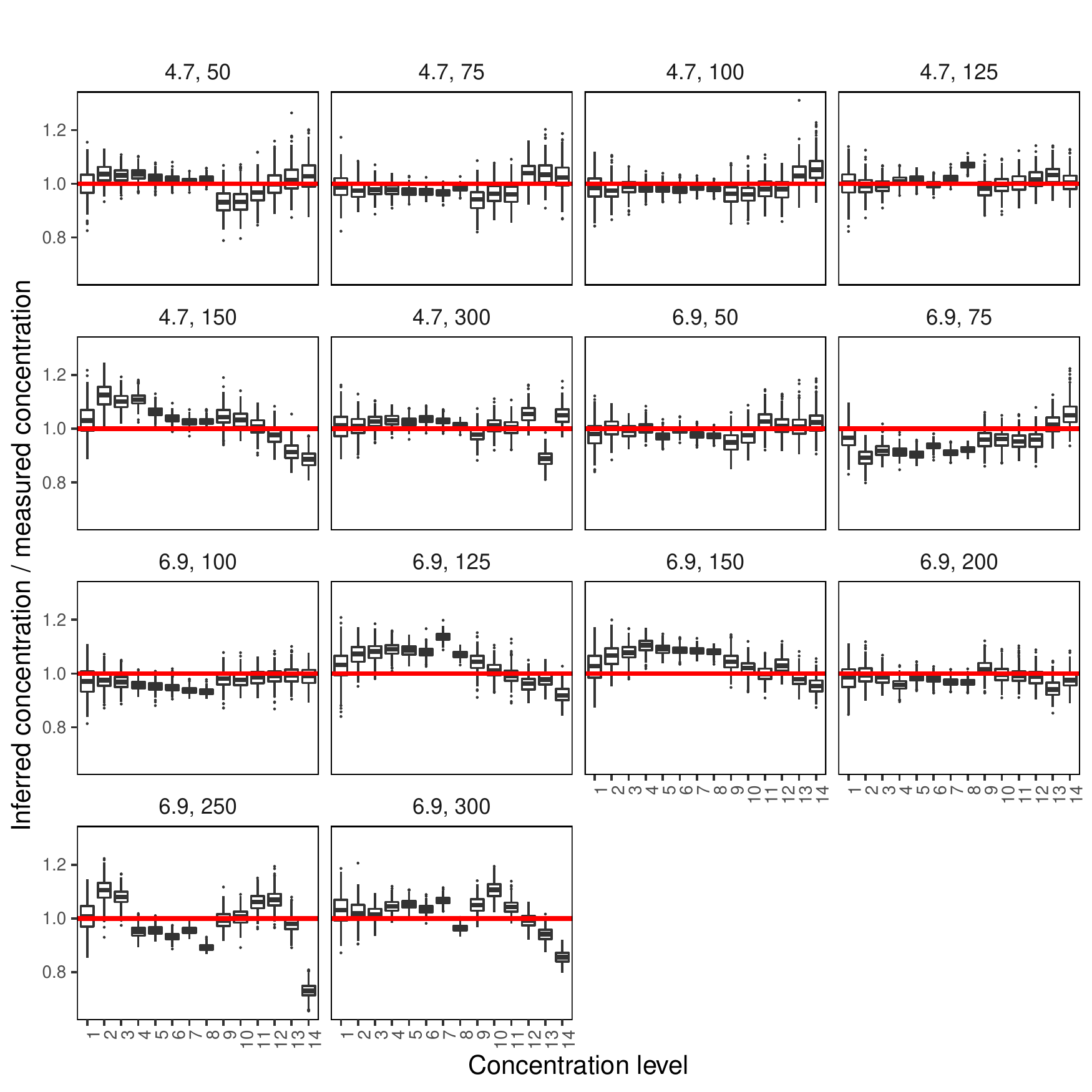}
    \caption{The ratio of inferred concentration to measured concentrations under each concentration level $i=1,\ldots,14$ and experiment condition (pH, NaCl concentration measured in mM).}
    \label{fig:lysozyme_model_check_con}
\end{figure}

Figure \ref{fig:lysozyme_model_check_con} shows the relationship between inferred concentrations and observed concentrations, and we see that the proposed method can effectively calibrate the concentration measurements with the information coming from LS and RI readings.
% (see, Figure \ref{fig:lysozyme_model_check_LS}, \ref{fig:lysozyme_model_check_RI}). 

%  $R$ and $\Delta n$
% Figure \ref{fig:lysozyme_model_check_LS}, \ref{fig:lysozyme_model_check_RI} and \ref{fig:lysozyme_model_check_con} present the results of posterior predictive checks on 

% \begin{figure}[H]
%     \centering
%     \includegraphics[width=\textwidth]{lysozyme_complete_pool_constant_dn/dc_noconmeas_log_normal_shorter_residuals_dRI.pdf}
%     \caption{Posterior predictive checks, relative difference in predicted and observed $\Delta n$, experiment condition (pH, NaCl). The reference line is colored in red.}
%     \label{fig:lysozyme_model_check_RI}
% \end{figure}

% \begin{figure}[H]
%     \centering
%     \includegraphics[width=\textwidth]{lysozyme_complete_pool_constant_dn/dc_noconmeas_log_normal_shorter_residuals_LS.pdf}
%     \caption{Posterior predictive checks, relative difference in predicted and observed Rayleigh ratio, experiment condition (pH, NaCl). The reference line is colored in red.}
%     \label{fig:lysozyme_model_check_LS}
% \end{figure}

% \rwm{This figure did not show up for me (not in the folder?) and the caption is unclear.}
%%%%%%%%%%

\section{Application to Aggregation Propensity Assessment in human $\gamma$S-crystallin}
\label{sec:gammaS_application}

In this section, we study the aggregation status of human $\gamma$S-crystallin (H$\gamma$S), a dominant structural component of the human eye lens.  H$\gamma$S is noteworthy for its ability to remain in solution at the extremely high concentrations necessary to give the lens its refractive power, while resisting aggregation; indeed, as the lens contains no mechanisms to either remove or replace aggregated H$\gamma$S, it must remain in solution for one's entire life \citep{Wistow:1988aa, Bloemendal:2004aa, Slingsby:2013aa}. Crystallin aggregation leads to cataract, the leading cause of blindness worldwide \citep{WHO}, and is hence of considerable scientific importance.  The transient oligomerization states of H$\gamma$S are poorly understood beyond dimers \citep{thorn2019structure}, and precise measurements of its $A_2$ values under different solution conditions are so far lacking, making it a natural target for investigation using SLS. 

\subsection{Experimental condition and data preparation}
DNA encoding the sequence of human $\gamma$S-crystallin (UniProt ID: CRYGS\_HUMAN) \citep{siezen1987human}, codon-optimized for expression in \textit{E. coli}, was purchased from Blue Heron (Bothell, WA). This gene was cloned into a pET28a(+) plasmid (Novagen, Darmstadt, Germany) containing an N-terminal 6$\times$ His tag and a TEV cleavage sequence (ENLFQG), which leaves a glycine in place of the initiator methionine. The protein was overexpressed in a Rosetta \textit{E. coli} cell line (DE3) using autoinduction as described by \citet{studier2005protein}. Cell pellets were collected via centrifugation at 4,000 rpm for 30 minutes, resuspended, lysed, and spun again at 14,000 rpm for 60 minutes. Finally, the protein was purified via nickel affinity chromatography, digested with TEV protease (produced in-house), and the His-tag removed using a nickel affinity chromatography step. Three experiments were conducted under the same solution condition (pH = 6.9, NaCl = 100mM); the experimental procedure is similar to that used for lysozyme.  

Table \ref{tb:gsc_LS_RI_measurements} shows the availability of RI and LS measurements under different nominal concentration levels, which is similar to that of lysozyme -- note that RI readings are not available for conditions with nominal concentrations $>20$ mg/mL, due to limitations of the refractometer for proteins of particularly high refractive index.

\begin{table}
\caption{\label{tb:gsc_LS_RI_measurements} LS and RI measurements under different nominal concentration levels within an experiment replicate under our experiment condition (pH = 6.9, NaCl concentration : 100mM). ``Y'' indicates the corresponding signal is included in the analysis, otherwise ``N''.}
% \centering
 \fbox{%
  \begin{tabular}{l | l l l l l l l l l l l l l l}
  \hline
nominal concentration level (i=) & 1 & 2 & 3 & 4 & 5 & 6 & 7 & 8 & 9 & 10 & 11 & 12 & 13 & 14 \\
nominal concentration (mg/mL) & 0.5 & 1 & 2 & 3 & 4 & 5 & 7.5 & 10 & 12.5 & 15 & 17.5 & 20 & 25 & 30 \\
   \hline
RI measurements & Y & Y & Y & Y & Y & Y & Y & Y & Y & Y & Y & Y & N & N \\
   \hline
LS measurements & Y & Y & Y & Y & Y & Y & Y & Y & Y & Y & Y & Y & Y & Y
   \end{tabular}}
\end{table}

The raw experimental data were cleaned before analysis, using the procedure described in Section \ref{sec:data_cleaning}.

\subsection{Model specification}
Although H$\gamma$S is generally assumed to be monomeric, it exists under very crowded conditions in the eye lens, where it avoids aggregation despite having mildly attractive intermolecular interactions \citep{delaye1983short}. The current understanding of the \emph{transient} oligomerization states of $\gamma$S-crystallin is limited; possibilities include both polydispersity and monodispersity with large, dynamically exchanging structures (scattering units).  With this in mind we consider the following candidate models for $M_{w}$ (here we omit the index $l$ as we only have one experimental condition, which was chosen to mimic the physiological situation):
\begin{itemize}
    \item $\mathcal{M}_{x}: M_{w} = x M$, $x = 1,2,\ldots,20$
    \item $\mathcal{M}_{21}: M_{w} \sim \mathcal{U}(M, 20M)$
    \item $\mathcal{M}_{22}: M_{w} \sim \mathcal{N}(\mu_{M_{w}}, \sigma^{2}_{M_{w}})$, where $\mu_{M_{w}} \sim \mathcal{U}(M, 20M)$, $\sigma^{2}_{M_{w}} \sim IG(1, (\frac{M}{3})^{2})$
\end{itemize}
\noindent where $M=20959.80$ g/mol. Models $\mathcal{M}_{x}, \ \ x = 1,2,\ldots,20$ assume monodispersity (with particle sizes ranging from 1 to 20 monomers), while $\mathcal{M}_{21}$ and $\mathcal{M}_{22}$ allow the co-existence of different aggregation states. In model $\mathcal{M}_{22}$, the prior for $\sigma^{2}_{M_{w}}$ is chosen to ensure that possible aggregates are close to the center.  Other hyperparameters for the H$\gamma$S models were chosen as per the lysozyme analysis (i.e., the same values were employed).

\subsection{Results}

\begin{figure}[htp]
    \centering
    \includegraphics[width=\textwidth]{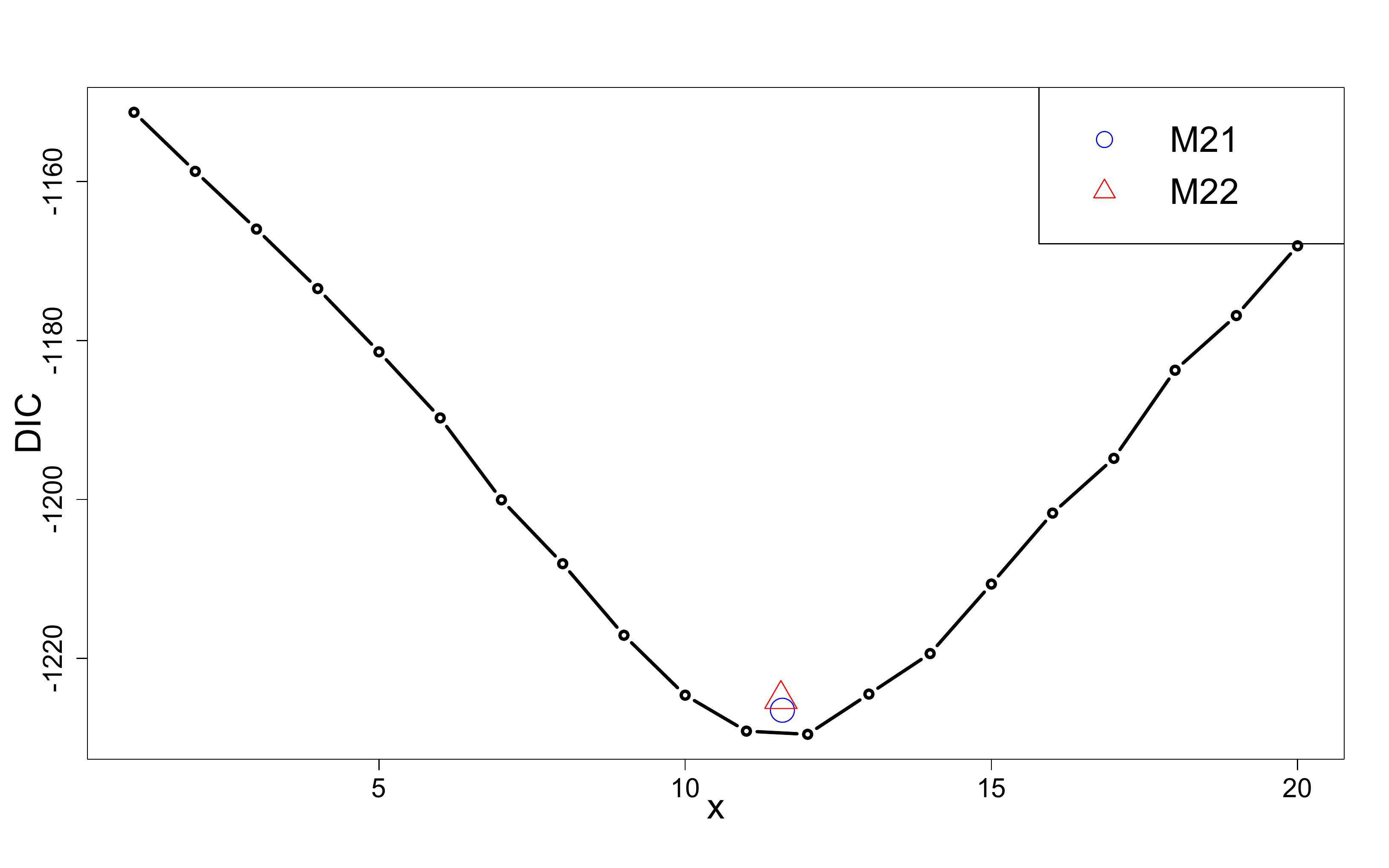}
    \caption{DIC values for $\mathcal{M}_{x}, \ \ x=1,2,\ldots,20$, $\mathcal{M}_{21}$ (blue circle) and $\mathcal{M}_{22}$ (red triangle). The x-axis values of points associated with  $\mathcal{M}_{21}$ and $\mathcal{M}_{22}$ are determined by the posterior mean of $M_{w}/M$.}
    \label{fig:DICxmer}
\end{figure}

\begin{figure}[htp]
    \centering
    \includegraphics[width=\textwidth]{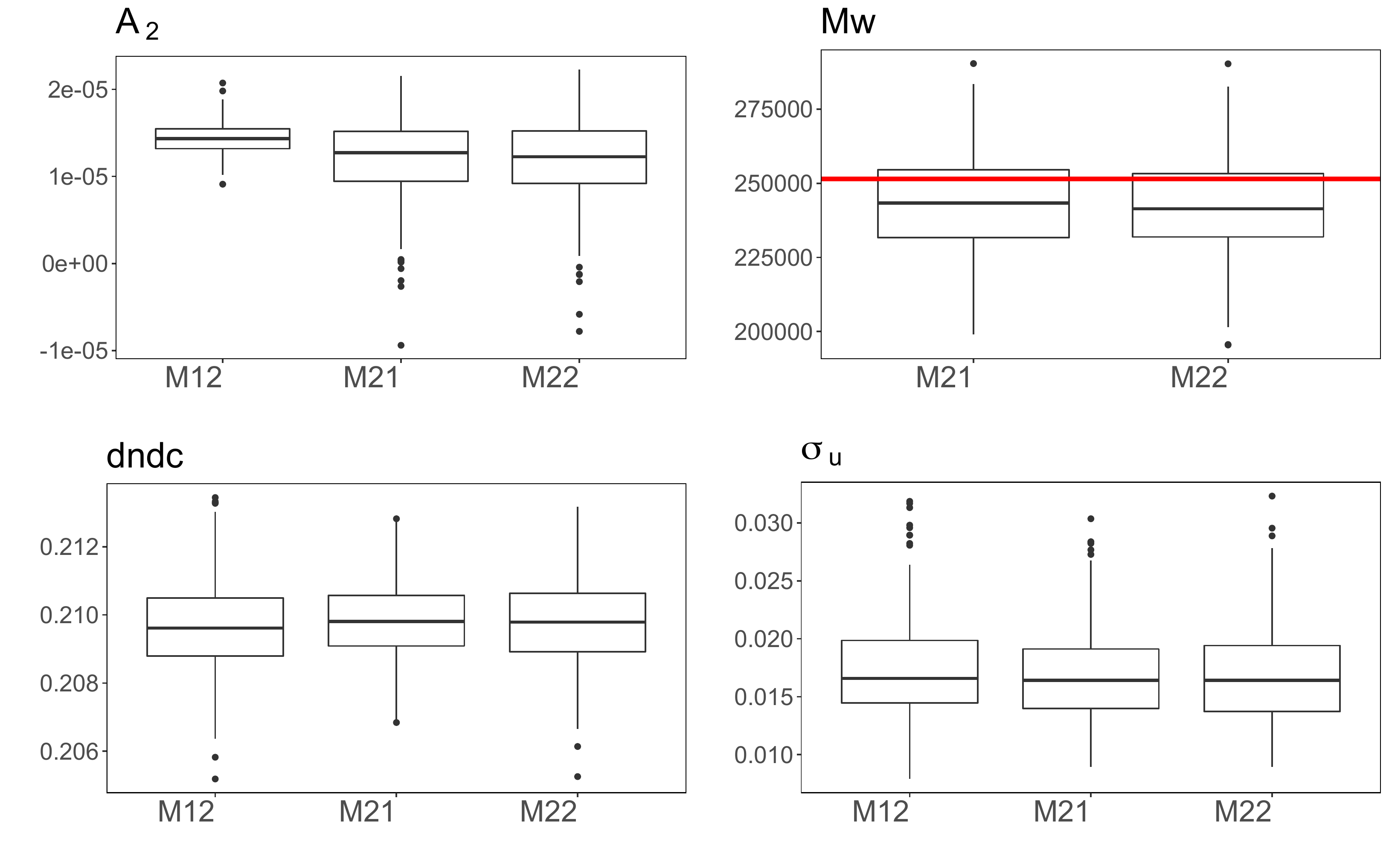}
    \caption{Boxplots of posterior samples of weight average molecular weight $M_{w}$ and $A_2$. The red horizontal line in the boxplot for $M_w$ indicates the value of $M_w$ under model $\mathcal{M}_{12}$, that is, $12 \times 20959.80 = 251517.6$ g/mol.}
    \label{fig:gsc_A2Mw}
\end{figure}

Figure \ref{fig:DICxmer} presents the DIC values for the candidate oligomerization state models. We observe that $x=12$ yields the smallest DIC value, which is similar to the DIC of $\mathcal{M}_{21}$ and $\mathcal{M}_{22}$. Figure \ref{fig:gsc_A2Mw} shows that $\mathcal{M}_{12}$, $\mathcal{M}_{21}$ and $\mathcal{M}_{22}$ yield similar posterior median mass estimates ($1.43$, $1.22$ and $1.29$ $\times 10^{-5}$ mL*mol/g$^2$, respectively) and probabilities of being positive ($1$, $0.988$ and $0.995$, respectively) for $A_2$, though the latter two models give wider posterior intervals. In addition, these models also yield similar inference on $M_{w}$, suggesting the dodecamer ($x=12$) might be the dominant structure in human $\gamma$S-crystallin solution under this solution condition (pH = 6.9, NaCl concentration: 100mM), with a nontrivial chance that decameric ($x=10$), undecameric ($x=11$) and tridecameric ($x=13$) forms might exist as well.  Taken together, the combination of large $M_{w}$ and positive $A_2$ suggests a ``self-avoiding cluster'' model for H$\gamma$S, in which monomers interact attractively to form moderately sized oligomers, with the oligomers tending to repel one another (possibly due to selective exposure of less favorable interaction sites on the surface of the cluster, with sites favorable to surface interaction occupied by interactions with other cluster members).

% \rwm{Lines are too thin, labels are too small.}

\subsubsection{Model assessment}

Table \ref{tb:gsc_LS_post_check} and \ref{tb:gsc_RI_post_check} show the posterior predictive $p$-values for models $\mathcal{M}_{21}$, $\mathcal{M}_{22}$ and $\mathcal{M}_{12}$, aggregated over replicates, for the Rayeligh ratio and $\Delta n$ readings. These $p$-values are close to $0.5$, which is indicative of a high degree of compatibility with the observed data.

\begin{table}
\caption{Posterior predictive $p$-values for Rayleigh ratio measurements under $\mathcal{M}_{21}$, $\mathcal{M}_{22}$ and $\mathcal{M}_{12}$.  \label{tb:gsc_LS_post_check}}
\centering
\fbox{\begin{tabular}{l|lll}
  \hline
  Concentration level (i=) & $\mathcal{M}_{21}$ & $\mathcal{M}_{22}$ & $\mathcal{M}_{12}$ \\ 
  \hline
  1 & 0.532 & 0.532 & 0.542 \\ 
  2 & 0.511 & 0.488 & 0.497 \\ 
  3 & 0.519 & 0.540 & 0.541 \\ 
  4 & 0.585 & 0.545 & 0.545 \\ 
  5 & 0.620 & 0.588 & 0.594 \\ 
  6 & 0.526 & 0.507 & 0.532 \\ 
  7 & 0.607 & 0.553 & 0.570 \\ 
  8 & 0.620 & 0.568 & 0.572 \\ 
  9 & 0.595 & 0.579 & 0.566 \\ 
  10 & 0.602 & 0.578 & 0.568 \\ 
  11 & 0.570 & 0.575 & 0.573 \\ 
  12 & 0.559 & 0.551 & 0.542 \\ 
  13 & 0.488 & 0.459 & 0.500 \\ 
  14 & 0.479 & 0.499 & 0.521 \\ 
   \hline
\end{tabular}}
\end{table}

\begin{table}
\caption{Posterior predictive $p$-values for $\Delta n$ measurements under $\mathcal{M}_{21}$, $\mathcal{M}_{22}$ and $\mathcal{M}_{12}$.  \label{tb:gsc_RI_post_check}}
\centering
\fbox{\begin{tabular}{l|lll}
  \hline
 Concentration level (i=) & $\mathcal{M}_{21}$ & $\mathcal{M}_{22}$ & $\mathcal{M}_{12}$ \\ 
  \hline
  1 & 0.356 & 0.376 & 0.374 \\ 
  2 & 0.352 & 0.349 & 0.355 \\ 
  3 & 0.393 & 0.386 & 0.371 \\ 
  4 & 0.357 & 0.358 & 0.316 \\ 
  5 & 0.427 & 0.418 & 0.412 \\ 
  6 & 0.469 & 0.463 & 0.473 \\ 
  7 & 0.473 & 0.484 & 0.479 \\ 
  8 & 0.623 & 0.612 & 0.603 \\ 
  9 & 0.518 & 0.552 & 0.550 \\ 
  10 & 0.579 & 0.583 & 0.573 \\ 
  11 & 0.444 & 0.438 & 0.430 \\ 
  12 & 0.389 & 0.393 & 0.376 \\ 
   \hline
\end{tabular}}
\end{table}

In passing, we note that when multiple competing models ($\mathcal{M}_{1}, \ldots, \mathcal{M}_{m}$) represent very different scientific mechanisms but yield similar performance in terms of the model selection criteria, Bayesian Model Averaging (BMA) \citep{hoeting1999bayesian} can be leveraged to incorporate the model uncertainty into the posterior inference of the physical quantities of interest (e.g., $A_2$ in this analysis), which is another advantage of pursuing a Bayesian solution to this problem. Given the MCMC samples, the marginal likelihood $P(\text{data}|\mathcal{M}_{j}), \ \ j=1,2,\ldots,m$ required in BMA can be calculated by various different methods, including one-block Metropolis–Hastings method \citep{chib2001marginal}, power posteriors and thermodynamic integration \citep{friel2008marginal,calderhead2009estimating}, etc.  Although we do not pursue this here (since the high-probability models in this case are already in strong substantive agreement), this approach would be viable in cases where greater differences were observed in posterior estimates.

% 0.5,1,2,3,4,5,7.5,10,12.5,15,17.5,20,25,30
\section{Simulation Study}
\label{sec:simulations}

A recurring theme in our analysis has been the importance of accounting for both uncertainty and measurement error in concentration.  Here, we conduct a systematic simulation study to shed light on the impact of sample size and error control in concentration on inferential accuracy, thereby providing guidance for the design of future experiments. As $A_{2}$ is the key physical quantity of interest in SLS experiments of the type discussed here, we focus on how well it can be estimated using the proposed Bayesian model. Specifically, the metrics for evaluating model performance are the bias of posterior means, the frequentist coverage and the width of posterior $95\%$ credible intervals. As there are many different components of the proposed model, some of which are subject to physical constraints or the precision level of instruments -- we fix those quantities at physically meaningful values and perform a full factorial design on the following four factors

\begin{enumerate}
    \item $A_2$ = $\pm 0.01, \pm 0.001, \pm 0.0001, \pm 0.00001$. These values cover the possible order of magnitude for $A_2$ in most real-world proteins. As chosen in the case studies, we consider the flat prior $\mathcal{N}(0,1^2)$ (with respect to the scales of possible values of $A_2$) across all simulation runs.
    \item $\sigma_{u}^{2} = (\log(1+0.01)/1.96)^2, (\log(1+0.05)/1.96)^2,
    (\log(1+0.10)/1.96)^2, (\log(1+0.20)/1.96)^2$. These choices correspond to the possible range of percentage errors in concentration measurements, $1\%$, $5\%$, $10\%$ and $20\%$, respectively.
    \item Prior on $\sigma_{u}^{2}$:
    \begin{itemize}
    \item Informative ($\sigma_{u}^{2} \sim IG(1+\frac{1}{2+2\sigma_{u}^{2}},\frac{1}{2})$): The informative prior is an Inverse-$\chi^{2}$ distribution with prior mean equal to corresponding true $\sigma_{u}^{2}$.
    \item Intermediate informative ($\sigma_{u}^{2} \sim IG(1,\sigma_{u}^{2})$): intermediate informative prior is an Inverse-Gamma distribution that concentrates a considerable amount of its mass around true value of $\sigma_{u}^{2}$ while being fairly spread.
    \item Weakly informative ($\sigma_{u}^{2} \sim IG(1, (\log(1+0.40)/1.96)^2)$):  the weakly informative prior is an Inverse-Gamma distribution that concentrates most of its mass around a realistic upper bound ($40\%$ relative error) of the measurement errors in concentrations while being fairly spread.
    \item No adjustment: Assuming the measured concentration is the true concentration. 
%  the somewhat informative prior is an Inverse-Gamma distribution that concentrates a considerable amount of its mass around true $\sigma_{u}^{2}$ while being fairly spread, the weakly informative prior is an Inverse-Gamma distribution that concentrates most of its mass around a realistic upper bound of the measurement errors in concentrations while being fairly spread.
    \end{itemize}
    \item Number of experiment replicates (dictating the sample size): 1, 2, 5, 10.
\end{enumerate}

We have a total of $8 \times 4 \times 4 \times 4 = 512$ settings, each of which is run for $100$ replicates. All MCMC chains are run for $300000$ iterations, and we store every $250$th iteration of the last $100000$ draws as posterior samples (the first $200000$ draws are discarded as burn-in). The data are generated using the model described in Section \ref{sec:Bayesian_model}, the concentration levels are set as those in Table \ref{tb:lysozyme_LS_RI_measurements}, and the true values of $\sigma^2_{R}$ and $\sigma^2_{\Delta n}$ are fixed at the posterior mean reported in Table \ref{fig:lysozyme_others_M1} to mimic the settings in real SLS experiments. Table \ref{tb:simulation_true_prior} shows the ground truth values and priors for parameters that are unchanged across different experimental runs. 

\begin{table}
\caption{\label{tb:simulation_true_prior} True values and priors for parameters that are unchanged across different experimental runs. The symbol ``-'' means no prior assigned to the corresponding quantity (i.e., known constant that is not inferred using the model).}
\centering
 \fbox{%
  \begin{tabular}{c| c | c}
  \hline
& True value &  Prior  \\ 
  \hline
$\sigma^2_{R}$  & $1 \times 10^{-11}$  & $IG(1, 10^{-10})$ \\
$\sigma^2_{\Delta n}$ & $2 \times 10^{-9}$ & $IG(1, 10^{-8})$ \\
$dn/dc$              & 0.20 &  $\mathcal{N}(0.195, 0.005^2)$ \\
% $A_2$               & -  & $\mathcal{N}(0,1^2)$ \\
$n_0$               & 1.33  & - \\
$\lambda$ & $657 \times 10^{-7}$ & - \\
$M$ & 14307 & - \\
   \hline
   \end{tabular}}
\end{table}

% ($\sigma^2_{R} = 1 \times 10^{-11}$ and $\sigma^2_{\Delta n} = 2 \times 10^{-9}$) and $n_0$ is fixed at $1.33$ 
% \begin{itemize}
%     \item $dn/dc \sim \mathcal{N}(0.195, 0.005^2)$
%     \item $\sigma^2_{R} \sim IG(1, 10^{-10})$
%     \item $\sigma^2_{\Delta n} \sim IG(1, 10^{-8})$
%     \item $A_2 \sim \mathcal{N}(0, 1^2)$
% \end{itemize}

% Figure \ref{fig:simulation_A2_10-2}, \ref{fig:simulation_A2_10-3}, \ref{fig:simulation_A2_10-4}, \ref{fig:simulation_A2_10-5} show the relative bias of posterior mean estimates of $A_2$. Figure \ref{fig:simulation_A2_10-2_coverage}, \ref{fig:simulation_A2_10-3_coverage}, \ref{fig:simulation_A2_10-4_coverage} and \ref{fig:simulation_A2_10-5_coverage} present the frequentist coverage probability of $95\%$ posterior intervals. Based on these results, we make the following observations and suggestions for future experiments:
% \rwm{figure labels in the previous paragraph do not compile properly.}

Table \ref{tb:simulation_relative_bias} and \ref{tb:simulation_coverage_prob} show the relative bias of posterior mean estimates and the frequentist coverage of $95\%$ posterior intervals, respectively. We have the following observations and suggestions for future experiments:

\begin{itemize}
    \item Estimating $A_2$ accurately becomes harder as it decreases, and improper treatment of the concentration error or small sample size can lead to substantial bias. 
    \item Compared to models where measurement errors in concentration are adjusted for, the model with no adjustment leads to larger relative bias when the error in concentrations is large ($>10\%$) and the absolute value of true $A_2$ is relatively large ($\geqslant 0.001$).
    \item The major concern with not adjusting for measurement errors in concentration is the ``no-adjustment'' model's inability to provide correct uncertainty estimates, that is, substantial undercoverage, given the error in concentration measurements are beyond $5\%$, and this problem still persists when the sample size is large.
    \item When the measurement errors are adjusted, upward biases are generally associated with negative $A_2$, while downward biases are mainly associated with positive $A_2$, regardless of other factors. It is also worth noting that we have slightly more difficulty in estimating negative $A_2$ values compared to positive $A_2$ values from a purely statistical standpoint, \emph{ceteris paribus}, though we are aware of the general perceptions that the detection of repulsive effects is harder than that of attractive ones in dilute solution from an experimental perspective. 
    
    \item Larger sample size improves performance (i.e., more replicates can help mitigate the bias), and we have diminishing returns going beyond 5 replicates. 
    \item A considerable gap exists between the performance of weakly informative prior and the other two prior choices, while the gap between two other priors is often minimal. For robustness, we recommend the ``intermediate informative'' prior.
    \item Larger discrepancies between measured concentrations and the actual concentrations can lead to substantial bias, and such bias may persist even if we know its magnitude very well and do many replicates, especially when the absolute value of $A_2$ is small. Therefore, high-precision in concentration measurements is crucial.  We suggest that experimenters use high-precision concentration measurement devices, reduce the errors in liquid handling, and more importantly, measure concentrations both before and after the LS and RI experiments, if possible.
\end{itemize}

\begin{table}
% \centering
\caption{\label{tb:simulation_relative_bias} Relative bias of posterior mean estimates of $A_{2}$ under different settings.}
\fbox{\resizebox{\textwidth}{!}{\begin{tabular}{lrrr|rrrr|rrrr|rrrr|rrrr}
\hline
& \multicolumn{3}{c}{True values} & \multicolumn{4}{c}{Informative} & \multicolumn{4}{c}{Intermediate informative} & \multicolumn{4}{c}{Weakly informative}  & \multicolumn{4}{c}{No adjustment}  \\
% & & True values & & & Informative &  &  &  & Intermediate &  & &  & Weakly & & &  & No adj. & &  \\ 
& $|A_{2}|$ & $A_2$ & concentration errors $\text{in}(\%)$ & 1 & 2 & 5 & 10 & 1 & 2 & 5 & 10 & 1 & 2 & 5 & 10 & 1 & 2 & 5 & 10 \\ 
\hline
 & 0.01 & 0.01 & 1.00 & 0.01 & 0.00 & 0.00 & 0.00 & 0.01 & 0.00 & 0.00 & 0.00 & -0.00 & -0.00 & -0.00 & -0.00 & 0.01 & 0.00 & 0.00 & -0.00 \\ 
 & 0.01 & 0.01 & 5.00 & 0.01 & 0.00 & 0.00 & 0.00 & 0.01 & 0.00 & 0.00 & 0.00 & -0.00 & -0.00 & -0.00 & -0.00 & 0.01 & 0.00 & 0.01 & 0.00 \\ 
 & 0.01 & 0.01 & 10.00 & 0.00 & 0.00 & 0.00 & 0.00 & 0.01 & 0.00 & 0.00 & 0.00 & -0.00 & -0.00 & -0.00 & -0.00 & 0.02 & 0.01 & 0.02 & 0.00 \\ 
   & 0.01 & 0.01 & 20.00 & 0.00 & -0.00 & 0.00 & -0.00 & 0.00 & -0.00 & 0.00 & -0.00 & -0.00 & -0.00 & -0.00 & -0.00 & 0.04 & 0.03 & 0.04 & 0.02 \\ 
   & 0.01 & -0.01 & 1.00 & 0.01 & 0.00 & 0.00 & -0.00 & 0.01 & 0.00 & 0.00 & -0.00 & 0.00 & 0.00 & 0.00 & -0.00 & 0.01 & 0.00 & 0.00 & -0.00 \\ 
   & 0.01 & -0.01 & 5.00 & 0.01 & 0.00 & 0.00 & -0.00 & 0.01 & 0.00 & 0.00 & -0.00 & 0.00 & 0.00 & 0.00 & -0.00 & 0.01 & 0.01 & 0.01 & 0.00 \\ 
   & 0.01 & -0.01 & 10.00 & 0.01 & 0.00 & 0.00 & -0.00 & 0.01 & 0.00 & 0.00 & -0.00 & 0.00 & 0.00 & 0.00 & -0.00 & 0.02 & 0.01 & 0.02 & 0.00 \\ 
   & 0.01 & -0.01 & 20.00 & 0.01 & 0.00 & 0.00 & -0.00 & 0.01 & 0.00 & 0.00 & -0.00 & 0.00 & 0.00 & 0.00 & -0.00 & 0.05 & 0.04 & 0.04 & 0.02 \\ 
  \hline
   & 0.001 & 0.001 & 1.00 & 0.00 & -0.00 & 0.00 & -0.00 & 0.00 & -0.00 & 0.00 & -0.00 & -0.01 & -0.01 & -0.00 & -0.00 & 0.00 & -0.00 & 0.00 & -0.00 \\ 
   & 0.001 & 0.001 & 5.00 & 0.00 & -0.00 & 0.00 & 0.00 & 0.00 & -0.00 & 0.00 & 0.00 & -0.01 & -0.01 & -0.00 & -0.00 & 0.00 & -0.00 & 0.00 & 0.00 \\ 
   & 0.001 & 0.001 & 10.00 & 0.00 & -0.00 & 0.00 & 0.00 & 0.00 & -0.00 & 0.00 & 0.00 & -0.01 & -0.01 & -0.00 & -0.00 & 0.00 & 0.00 & 0.01 & 0.00 \\ 
   & 0.001 & 0.001 & 20.00 & -0.00 & -0.01 & 0.00 & -0.00 & -0.00 & -0.01 & 0.00 & -0.00 & -0.01 & -0.01 & -0.00 & -0.00 & 0.01 & 0.01 & 0.02 & 0.01 \\ 
   & 0.001 & -0.001 & 1.00 & 0.01 & 0.01 & 0.00 & 0.00 & 0.01 & 0.01 & 0.00 & 0.00 & 0.04 & 0.03 & 0.01 & -0.00 & 0.01 & 0.01 & 0.00 & 0.00 \\ 
   & 0.001 & -0.001 & 5.00 & 0.02 & 0.02 & 0.01 & 0.00 & 0.02 & 0.02 & 0.01 & 0.00 & 0.04 & 0.03 & 0.01 & -0.00 & 0.02 & 0.01 & 0.01 & 0.00 \\ 
   & 0.001 & -0.001 & 10.00 & 0.04 & 0.03 & 0.02 & 0.00 & 0.04 & 0.03 & 0.02 & 0.00 & 0.04 & 0.04 & 0.01 & -0.00 & 0.03 & 0.02 & 0.03 & 0.00 \\ 
   & 0.001 & -0.001 & 20.00 & 0.05 & 0.04 & 0.02 & 0.00 & 0.05 & 0.04 & 0.02 & 0.00 & 0.04 & 0.04 & 0.02 & 0.00 & 0.07 & 0.05 & 0.06 & 0.02 \\ 
  \hline
   & 0.0001 & 0.0001 & 1.00 & -0.04 & -0.03 & -0.00 & -0.00 & -0.05 & -0.04 & -0.00 & -0.00 & -0.18 & -0.07 & 0.06 & 0.04 & -0.04 & -0.03 & -0.00 & -0.00 \\ 
   & 0.0001 & 0.0001 & 5.00 & -0.10 & -0.06 & -0.03 & -0.01 & -0.09 & -0.06 & -0.03 & -0.01 & -0.17 & -0.08 & 0.04 & 0.04 & -0.06 & -0.04 & -0.02 & -0.00 \\ 
   & 0.0001 & 0.0001 & 10.00 & -0.17 & -0.11 & -0.07 & -0.02 & -0.16 & -0.10 & -0.07 & -0.02 & -0.18 & -0.10 & 0.00 & 0.03 & -0.10 & -0.07 & -0.05 & -0.00 \\ 
   & 0.0001 & 0.0001 & 20.00 & -0.28 & -0.21 & -0.15 & -0.06 & -0.27 & -0.20 & -0.16 & -0.06 & -0.22 & -0.14 & -0.08 & -0.01 & -0.19 & -0.13 & -0.11 & -0.01 \\ 
   & 0.0001 & -0.0001 & 1.00 & 0.06 & 0.04 & 0.01 & 0.00 & 0.06 & 0.04 & 0.01 & 0.00 & 0.25 & 0.17 & -0.02 & -0.04 & 0.05 & 0.04 & 0.01 & 0.00 \\ 
   & 0.0001 & -0.0001 & 5.00 & 0.12 & 0.08 & 0.04 & 0.01 & 0.12 & 0.07 & 0.04 & 0.01 & 0.25 & 0.18 & 0.00 & -0.04 & 0.08 & 0.05 & 0.04 & 0.00 \\ 
   & 0.0001 & -0.0001 & 10.00 & 0.23 & 0.16 & 0.10 & 0.02 & 0.21 & 0.15 & 0.10 & 0.03 & 0.26 & 0.20 & 0.05 & -0.02 & 0.14 & 0.09 & 0.09 & 0.01 \\ 
   & 0.0001 & -0.0001 & 20.00 & 0.37 & 0.29 & 0.21 & 0.07 & 0.36 & 0.29 & 0.21 & 0.07 & 0.29 & 0.26 & 0.15 & 0.03 & 0.29 & 0.20 & 0.21 & 0.04 \\ 
  \hline
   & 0.00001 & 0.00001 & 1.00 & -0.49 & -0.36 & -0.06 & -0.03 & -0.51 & -0.37 & -0.06 & -0.03 & -2.15 & -1.17 & 0.42 & 0.46 & -0.47 & -0.35 & -0.05 & -0.03 \\ 
   & 0.00001 & 0.00001 & 5.00 & -1.09 & -0.70 & -0.35 & -0.08 & -1.04 & -0.65 & -0.34 & -0.07 & -2.14 & -1.23 & 0.19 & 0.42 & -0.67 & -0.47 & -0.29 & -0.01 \\ 
   & 0.00001 & 0.00001 & 10.00 & -1.99 & -1.30 & -0.84 & -0.23 & -1.86 & -1.22 & -0.85 & -0.24 & -2.19 & -1.47 & -0.21 & 0.27 & -1.17 & -0.76 & -0.65 & -0.04 \\ 
   & 0.00001 & 0.00001 & 20.00 & -3.25 & -2.48 & -1.84 & -0.65 & -3.16 & -2.44 & -1.89 & -0.66 & -2.53 & -2.00 & -1.20 & -0.19 & -2.35 & -1.59 & -1.53 & -0.21 \\ 
   & 0.00001 & -0.00001 & 1.00 & 0.51 & 0.37 & 0.06 & 0.03 & 0.53 & 0.38 & 0.06 & 0.03 & 2.20 & 1.33 & -0.38 & -0.46 & 0.47 & 0.35 & 0.06 & 0.03 \\ 
   & 0.00001 & -0.00001 & 5.00 & 1.12 & 0.71 & 0.37 & 0.08 & 1.05 & 0.66 & 0.36 & 0.08 & 2.25 & 1.36 & -0.13 & -0.41 & 0.68 & 0.47 & 0.30 & 0.01 \\ 
   & 0.00001 & -0.00001 & 10.00 & 2.05 & 1.35 & 0.87 & 0.24 & 1.91 & 1.24 & 0.89 & 0.25 & 2.30 & 1.53 & 0.28 & -0.26 & 1.23 & 0.79 & 0.69 & 0.05 \\ 
   & 0.00001 & -0.00001 & 20.00 & 3.32 & 2.55 & 1.90 & 0.66 & 3.24 & 2.51 & 1.93 & 0.68 & 2.56 & 2.12 & 1.25 & 0.20 & 2.45 & 1.68 & 1.64 & 0.26 \\ 
   \hline
   \hline
  \hline
\end{tabular}}}
\end{table}

\begin{table}
% \centering
\caption{\label{tb:simulation_coverage_prob} Coverage rates of $95\%$ credible intervals for $A_{2}$ under different settings.}
\fbox{\resizebox{\textwidth}{!}{\begin{tabular}{lrrr|rrrr|rrrr|rrrr|rrrr}
\hline
& \multicolumn{3}{c}{True values} & \multicolumn{4}{c}{Informative} & \multicolumn{4}{c}{Intermediate informative} & \multicolumn{4}{c}{Weakly informative}  & \multicolumn{4}{c}{No adjustment}  \\
% & & True values  & &  & Informative & & & & Intermediate &  & &  & Weakly & & &  & No adj. & &  \\ 
& $|A_{2}|$ & $A_2$ & concentration errors $\text{in}(\%)$ & 1 & 2 & 5 & 10 & 1 & 2 & 5 & 10 & 1 & 2 & 5 & 10 & 1 & 2 & 5 & 10 \\ 
  \hline
 & 0.01 & 0.01 & 1.00 & 1.00 & 0.99 & 0.95 & 0.94 & 1.00 & 0.99 & 0.95 & 0.96 & 1.00 & 1.00 & 0.96 & 0.95 & 1.00 & 0.99 & 0.93 & 0.98 \\ 
   & 0.01 & 0.01 & 5.00 & 1.00 & 0.98 & 0.97 & 0.96 & 1.00 & 0.99 & 0.97 & 0.95 & 1.00 & 1.00 & 0.97 & 0.97 & 0.80 & 0.79 & 0.82 & 0.88 \\ 
   & 0.01 & 0.01 & 10.00 & 1.00 & 1.00 & 0.97 & 0.94 & 1.00 & 1.00 & 0.97 & 0.94 & 1.00 & 1.00 & 0.97 & 0.93 & 0.72 & 0.74 & 0.76 & 0.83 \\ 
   & 0.01 & 0.01 & 20.00 & 0.99 & 1.00 & 0.97 & 0.93 & 0.99 & 1.00 & 0.97 & 0.94 & 1.00 & 0.99 & 0.97 & 0.94 & 0.66 & 0.68 & 0.66 & 0.77 \\ 
   & 0.01 & -0.01 & 1.00 & 1.00 & 1.00 & 0.95 & 0.99 & 1.00 & 0.99 & 0.95 & 0.98 & 1.00 & 1.00 & 0.95 & 0.98 & 0.98 & 0.99 & 0.93 & 0.97 \\ 
   & 0.01 & -0.01 & 5.00 & 1.00 & 0.99 & 0.94 & 0.99 & 1.00 & 1.00 & 0.95 & 0.98 & 1.00 & 1.00 & 0.94 & 0.99 & 0.77 & 0.74 & 0.81 & 0.88 \\ 
   & 0.01 & -0.01 & 10.00 & 1.00 & 1.00 & 0.94 & 0.99 & 1.00 & 0.99 & 0.94 & 0.99 & 1.00 & 1.00 & 0.95 & 0.99 & 0.72 & 0.73 & 0.76 & 0.83 \\ 
   & 0.01 & -0.01 & 20.00 & 1.00 & 1.00 & 0.93 & 0.97 & 1.00 & 1.00 & 0.92 & 0.97 & 1.00 & 1.00 & 0.94 & 0.98 & 0.65 & 0.69 & 0.67 & 0.78 \\ 
  \hline
   & 0.001 & 0.001 & 1.00 & 0.99 & 0.97 & 0.95 & 0.94 & 1.00 & 0.98 & 0.93 & 0.93 & 1.00 & 1.00 & 1.00 & 0.99 & 0.99 & 0.98 & 0.92 & 0.90 \\ 
   & 0.001 & 0.001 & 5.00 & 0.98 & 0.99 & 0.93 & 0.93 & 0.98 & 0.97 & 0.94 & 0.93 & 1.00 & 1.00 & 1.00 & 1.00 & 0.92 & 0.85 & 0.72 & 0.76 \\ 
   & 0.001 & 0.001 & 10.00 & 0.97 & 0.98 & 0.94 & 0.94 & 0.97 & 0.96 & 0.93 & 0.95 & 1.00 & 1.00 & 1.00 & 0.98 & 0.77 & 0.67 & 0.62 & 0.70 \\ 
   & 0.001 & 0.001 & 20.00 & 0.97 & 0.94 & 0.93 & 0.95 & 0.95 & 0.92 & 0.94 & 0.95 & 1.00 & 1.00 & 0.98 & 0.96 & 0.65 & 0.62 & 0.51 & 0.60 \\ 
   & 0.001 & -0.001 & 1.00 & 1.00 & 0.99 & 0.95 & 0.99 & 1.00 & 0.98 & 0.96 & 0.99 & 1.00 & 1.00 & 0.99 & 1.00 & 0.99 & 0.95 & 0.95 & 0.96 \\ 
   & 0.001 & -0.001 & 5.00 & 1.00 & 0.99 & 0.92 & 0.98 & 0.98 & 0.97 & 0.90 & 0.98 & 1.00 & 1.00 & 1.00 & 1.00 & 0.81 & 0.75 & 0.82 & 0.85 \\ 
   & 0.001 & -0.001 & 10.00 & 1.00 & 0.99 & 0.92 & 0.98 & 0.98 & 0.98 & 0.91 & 0.98 & 1.00 & 1.00 & 0.98 & 0.99 & 0.70 & 0.72 & 0.77 & 0.83 \\ 
   & 0.001 & -0.001 & 20.00 & 1.00 & 0.99 & 0.93 & 0.99 & 0.99 & 0.98 & 0.93 & 0.99 & 1.00 & 1.00 & 0.96 & 0.99 & 0.69 & 0.66 & 0.73 & 0.80 \\ 
  \hline
   & 0.0001 & 0.0001 & 1.00 & 1.00 & 0.98 & 0.97 & 0.97 & 1.00 & 0.98 & 0.97 & 0.96 & 1.00 & 1.00 & 1.00 & 1.00 & 1.00 & 0.98 & 0.97 & 0.95 \\ 
   & 0.0001 & 0.0001 & 5.00 & 0.99 & 0.98 & 0.94 & 0.99 & 0.99 & 0.95 & 0.94 & 0.96 & 1.00 & 1.00 & 1.00 & 1.00 & 0.97 & 0.89 & 0.89 & 0.95 \\ 
   & 0.0001 & 0.0001 & 10.00 & 0.98 & 0.95 & 0.92 & 0.98 & 0.97 & 0.91 & 0.91 & 0.97 & 1.00 & 1.00 & 1.00 & 1.00 & 0.85 & 0.81 & 0.85 & 0.87 \\ 
   & 0.0001 & 0.0001 & 20.00 & 0.99 & 0.94 & 0.91 & 0.96 & 0.97 & 0.90 & 0.88 & 0.96 & 1.00 & 1.00 & 0.99 & 0.99 & 0.75 & 0.75 & 0.81 & 0.84 \\ 
   & 0.0001 & -0.0001 & 1.00 & 1.00 & 0.98 & 0.98 & 0.97 & 1.00 & 0.98 & 0.98 & 0.96 & 1.00 & 1.00 & 1.00 & 1.00 & 1.00 & 0.97 & 0.96 & 0.96 \\ 
   & 0.0001 & -0.0001 & 5.00 & 0.99 & 0.98 & 0.93 & 0.98 & 0.99 & 0.95 & 0.92 & 0.97 & 1.00 & 1.00 & 1.00 & 1.00 & 0.92 & 0.83 & 0.88 & 0.89 \\ 
   & 0.0001 & -0.0001 & 10.00 & 0.99 & 0.95 & 0.91 & 0.98 & 0.97 & 0.91 & 0.90 & 0.98 & 1.00 & 1.00 & 1.00 & 1.00 & 0.77 & 0.76 & 0.84 & 0.84 \\ 
   & 0.0001 & -0.0001 & 20.00 & 0.97 & 0.93 & 0.89 & 0.96 & 0.95 & 0.93 & 0.89 & 0.96 & 1.00 & 1.00 & 0.98 & 1.00 & 0.72 & 0.71 & 0.77 & 0.83 \\ 
  \hline
   & 0.00001 & 0.00001 & 1.00 & 1.00 & 0.98 & 0.97 & 0.97 & 1.00 & 0.98 & 0.98 & 0.96 & 1.00 & 1.00 & 1.00 & 1.00 & 1.00 & 0.97 & 0.97 & 0.96 \\ 
   & 0.00001 & 0.00001 & 5.00 & 1.00 & 0.99 & 0.94 & 0.98 & 0.99 & 0.94 & 0.93 & 0.96 & 1.00 & 1.00 & 1.00 & 1.00 & 0.94 & 0.85 & 0.90 & 0.93 \\ 
   & 0.00001 & 0.00001 & 10.00 & 0.99 & 0.94 & 0.89 & 0.97 & 0.97 & 0.92 & 0.87 & 0.97 & 1.00 & 1.00 & 1.00 & 1.00 & 0.82 & 0.78 & 0.84 & 0.84 \\ 
   & 0.00001 & 0.00001 & 20.00 & 0.99 & 0.92 & 0.90 & 0.96 & 0.98 & 0.93 & 0.91 & 0.96 & 1.00 & 1.00 & 0.99 & 0.99 & 0.75 & 0.72 & 0.76 & 0.84 \\ 
   & 0.00001 & -0.00001 & 1.00 & 1.00 & 0.98 & 0.98 & 0.96 & 1.00 & 0.98 & 0.97 & 0.97 & 1.00 & 1.00 & 1.00 & 1.00 & 1.00 & 0.98 & 0.97 & 0.97 \\ 
   & 0.00001 & -0.00001 & 5.00 & 1.00 & 0.99 & 0.94 & 0.99 & 1.00 & 0.97 & 0.93 & 0.97 & 1.00 & 1.00 & 1.00 & 1.00 & 0.92 & 0.83 & 0.87 & 0.91 \\ 
   & 0.00001 & -0.00001 & 10.00 & 0.99 & 0.94 & 0.90 & 0.98 & 0.96 & 0.93 & 0.88 & 0.97 & 1.00 & 1.00 & 1.00 & 1.00 & 0.80 & 0.79 & 0.85 & 0.86 \\ 
   & 0.00001 & -0.00001 & 20.00 & 0.97 & 0.94 & 0.90 & 0.97 & 0.96 & 0.92 & 0.90 & 0.96 & 1.00 & 1.00 & 0.99 & 1.00 & 0.74 & 0.72 & 0.76 & 0.82 \\ 
   \hline
\end{tabular}}}
\end{table}
%%%% dn/dc
% \begin{figure}[H]
%     \centering
%     \includegraphics[width=\textwidth]{dn/dc_relative_bias.pdf}
%     \caption{Relative Bias for $dn/dc$, true $dn/dc = 0.2$ mL/g. The reference line is colored in red.}
%     \label{fig:simulation_dn/dc_relative_bias}
% \end{figure}

% \begin{figure}[H]
%     \centering
%     \includegraphics[width=\textwidth]{dn/dc_coverage.pdf}
%     \caption{Coverage probability of $95\%$ credible intervals for $dn/dc$, true $dn/dc = 0.20$ mL/g. The reference line is colored in red.}
%     \label{fig:simulation_dn/dc_coverage}
% \end{figure}

\section{Discussion}
\label{sec:discussion}

\subsection{Considerations for Experimental Procedure}
As illustrated in Section \ref{sec:simulations}, controlling errors in concentrations is the key to accurate static light scattering experiments. Naive methods with no adjustment for concentration errors can lead to reliable estimates when the relative error in concentration is extremely small (e.g. $1\%$). As the concentration errors become larger, we at least need to have some knowledge about the possible range of the measurement errors in concentrations and account for such errors in the model, and more importantly, larger sample sizes (e.g., more experimental replicates) are required to mitigate the bias caused by the concentration errors.  However, good results can be obtained with non-vanishing levels of concentration error, so long as an appropriate error model is used and adequate numbers of replicates are performed.  While error reduction via improved procedure is always a priority, building in replicates and avoiding ``no-adjustment'' models is recommended in practice.  

%CTB: These are cool things that I am commenting out for this paper
%\subsection{Angle-dependent Effects for Large Particles}
%
%\begin{itemize}
%\item We can handle $r_g$ and $P(\theta)$, probably, so long as $r_g>\lambda/20$ or so.  (Should say what sorts of stuff are that big.)  Perhaps show more complex model?
%\end{itemize}
%
%\subsection{Refinements at High Concentration}
%
%\begin{itemize}
%\item $A_3$ and friends go here.
%\end{itemize}

\section{Conclusion}
\label{sec:conclusion}

In this article, we proposed a novel Bayesian model for static light scattering (SLS) data that is sufficiently flexible to accommodate the complex relationship between physical quantities and their measurements, and to account for measurement errors. We ran simulation studies to gain insights about how measurement errors and sample size can affect the estimation of the second virial coefficient, and converted these insights into actionable guidance for future experiments. With the proposed model, we studied the protein aggregation behavior of two important proteins, lysozyme and human $\gamma$S-crystallin, in the former case identifying the conditions under which monomers transition from repulsive to attractive interactions, and in the second case showing the presence of a distinctive ``self-avoiding cluster'' structure in which monomers form oligomers of approximately dodecameric order which then interact repulsively.  Facilitating this was a protocol for cleaning and pre-processing SLS data, which provides a largely automated way to remove common artifacts and detect problems in data acquisition.

This article demonstrates the great value of Bayesian statistics in advancing data analysis within the biophysical context. Firstly, Bayesian analysis provides a principled way to update our beliefs about physical quantities using a combination of existing knowledge and experimental data. Secondly, though error modeling from a frequentist perspective is powerful, it can suffer from identitifiability problems if the error mechanism is not precisely known, or if certain classes of errors cannot be strictly ruled out. In contrast, Bayesian treatments are less sensitive to such difficulties so long as the posteriors remain characterizable, and informative priors can aid in filling in information that the data alone cannot supply.  As considerable background information is often possible in biophysical settings, this is a natural context for employing informative Bayesian analysis.  Thirdly, the Bayesian perspective can provide fully probabilistic answers to many scientific questions of interest, e.g, questions such as ``what is the probability of $A_2$ being positive given the experimental data?''  This advantage is highly valuable for problems such as $A_2$ estimation, where measurement is inherently difficult and residual uncertainty is expected to be large. Last but not least, continued advances in computational techniques mean that the ``Bayesian crank'' can be easily implemented using various freely available software packages, making it easier to supply solutions to practitioners without requiring them to be experts in e.g. MCMC simulation.

Given a powerful and flexible statistical model for the analysis of static light scattering data, researchers will be able to gain better understanding of the mechanisms governing protein aggregation. Such advances have the potential to inform areas such as medical research to develop better treatments for diseases such as Alzheimer’s and Parkinson’s Diseases, which are caused by protein aggregation. 

In closing, we comment on four potential directions for future work. First of all, we only work with the LS readings from angle $\theta = 90^{\circ}$ in this analysis.  Incorporating additional angles where available may improve precision, although it then becomes necessary to account for additional sources of error associated with mechanisms such as differences in detector alignment or sensitivity.  Secondly, this work is concerned with small proteins with $P(r_{g}, \theta) \approx 1$, and it is natural to consider extending our approach to large particles. Such an extension also requires further investigation on the use of readings from angles other than $\theta = 90^{\circ}$. Thirdly, motivated by the need to inform simulation-based work on protein aggregation, it would be interesting to consider whether higher-order virial coefficients could be inferred.  While present experimental methodology lacks the precision required for such analyses in settings like those studied here, future developments may remove this barrier.  Finally, the concentration levels in the experiments analyzed in this paper are chosen based on the experimenters' heuristics and the difficulty posed by different concentration ranges for sample preparation. It seems natural to attempt to improve on this by setting concentrations using sampling design theories for regression models \citep[see, e.g.,][]{elfving1952optimum,dette1993elfving,dette1996note,gilmour2012optimum}, potentially leading to more efficient experiments with similar inferential power. 

% multiple angles
% large particles
% third virial coefficient
% experimental design

% \begin{itemize}
% \item Summarize what has been said.
% \item Our shit is awesome.
% \item Now the world is a different place - Alzheimer's is going to be cured, and our awesome modeling scheme is totally the reason.
% \item Many extensions are possible, and should be done by someone.
% \item Philosophical moment: this work highlights the importance of close coupling between statistical modeling and experiment.  Bayesian methods offer a powerful and flexible way of incorporating the experimenter's prior knowledge, as well as leveraging multiple sources of information in a systematic way.  At the same time, ``old fashioned'' data analytic methods also have an important place in helping to eliminate uninteresting but possibly significant sources of error with a minimum of intervention (allowing the analyst to focus on the mechanisms of key importance).  By bringing modern statistics to bear on classic experiments, we can greatly expand the scope of what can be achieved.
% \end{itemize}
\section{Supplementary material}
R and JAGS codes along with the data for the computations in this paper are available from \url{https://github.com/fyin-stats/bayes-light-scattering}.

\section*{Acknowledgments}
The authors thank Dr. Dmitry Fishman, Director of the UCI Laser Spectroscopy Labs, Dr. Gianmarc Grazioli, who was a post-doctoral researcher at NCASD lab at UCI and is currently an assistant Professor at San Jos\'e State University, for insightful suggestions about this work and Andrew Meyer and Wyatt Instruments for assistance with the light-scattering instrument. This research was supported in part by NSF award DMS-1361425 to C.T.B. and R.W.M. and NIH awards S10OD021594 and 2R01EY021514 to R.W.M. 

% \FloatBarrier\clearpage
\section*{Appendix A: Table for relative width (average width / true $|A_2|$ value) of $95\%$ posterior credible intervals of $A_2$}
\begin{table}
\caption{Relative width (average width / true $|A_2|$ value) of $95\%$ posterior credible intervals of $A_2$ under different settings. \label{tb:simulation_relative_width}}
\centering
\fbox{\resizebox{\textwidth}{!}{\begin{tabular}{lrrr|rrrr|rrrr|rrrr|rrrr}
\hline
& \multicolumn{3}{c}{True values} & \multicolumn{4}{c}{Informative} & \multicolumn{4}{c}{Intermediate informative} & \multicolumn{4}{c}{Weakly informative}  & \multicolumn{4}{c}{No adjustment}  \\
% & & True values  & &  & Informative & & & & Intermediate &  & &  & Weakly & & &  & No adj. & &  \\ 
& $|A_{2}|$ & $A_2$ & concentration errors $\text{in}(\%)$ & 1 & 2 & 5 & 10 & 1 & 2 & 5 & 10 & 1 & 2 & 5 & 10 & 1 & 2 & 5 & 10 \\ 
  \hline
 & 0.01 & 0.01 & 1.00 & 0.06 & 0.04 & 0.02 & 0.01 & 0.06 & 0.04 & 0.02 & 0.01 & 0.09 & 0.05 & 0.02 & 0.02 & 0.07 & 0.04 & 0.02 & 0.02 \\ 
   & 0.01 & 0.01 & 5.00 & 0.07 & 0.04 & 0.02 & 0.02 & 0.07 & 0.05 & 0.02 & 0.02 & 0.09 & 0.05 & 0.02 & 0.02 & 0.09 & 0.07 & 0.04 & 0.03 \\ 
   & 0.01 & 0.01 & 10.00 & 0.08 & 0.05 & 0.02 & 0.02 & 0.08 & 0.05 & 0.02 & 0.02 & 0.09 & 0.05 & 0.03 & 0.02 & 0.14 & 0.11 & 0.07 & 0.05 \\ 
  & 0.01 & 0.01 & 20.00 & 0.09 & 0.05 & 0.03 & 0.02 & 0.09 & 0.05 & 0.03 & 0.02 & 0.09 & 0.05 & 0.03 & 0.02 & 0.22 & 0.18 & 0.13 & 0.09 \\ 
   & 0.01 & -0.01 & 1.00 & 0.08 & 0.05 & 0.03 & 0.02 & 0.08 & 0.05 & 0.03 & 0.02 & 0.11 & 0.06 & 0.03 & 0.02 & 0.08 & 0.05 & 0.03 & 0.02 \\ 
   & 0.01 & -0.01 & 5.00 & 0.10 & 0.06 & 0.03 & 0.02 & 0.10 & 0.06 & 0.03 & 0.02 & 0.11 & 0.06 & 0.03 & 0.02 & 0.11 & 0.08 & 0.05 & 0.03 \\ 
   & 0.01 & -0.01 & 10.00 & 0.11 & 0.06 & 0.03 & 0.02 & 0.11 & 0.06 & 0.03 & 0.02 & 0.11 & 0.06 & 0.03 & 0.02 & 0.16 & 0.12 & 0.08 & 0.06 \\ 
   & 0.01 & -0.01 & 20.00 & 0.11 & 0.07 & 0.03 & 0.02 & 0.11 & 0.07 & 0.03 & 0.02 & 0.11 & 0.06 & 0.03 & 0.02 & 0.25 & 0.20 & 0.15 & 0.11 \\ 
  \hline
   & 0.001 & 0.001 & 1.00 & 0.05 & 0.03 & 0.02 & 0.01 & 0.05 & 0.03 & 0.02 & 0.01 & 0.24 & 0.13 & 0.06 & 0.03 & 0.05 & 0.03 & 0.02 & 0.01 \\ 
   & 0.001 & 0.001 & 5.00 & 0.08 & 0.05 & 0.03 & 0.02 & 0.08 & 0.05 & 0.03 & 0.02 & 0.24 & 0.13 & 0.06 & 0.03 & 0.06 & 0.04 & 0.02 & 0.01 \\ 
   & 0.001 & 0.001 & 10.00 & 0.11 & 0.07 & 0.04 & 0.03 & 0.11 & 0.07 & 0.04 & 0.03 & 0.24 & 0.14 & 0.07 & 0.04 & 0.07 & 0.05 & 0.03 & 0.02 \\ 
   & 0.001 & 0.001 & 20.00 & 0.16 & 0.11 & 0.06 & 0.05 & 0.16 & 0.10 & 0.06 & 0.04 & 0.25 & 0.15 & 0.08 & 0.05 & 0.11 & 0.08 & 0.06 & 0.04 \\ 
   & 0.001 & -0.001 & 1.00 & 0.14 & 0.09 & 0.05 & 0.03 & 0.14 & 0.09 & 0.05 & 0.03 & 0.46 & 0.29 & 0.16 & 0.09 & 0.13 & 0.08 & 0.05 & 0.03 \\ 
   & 0.001 & -0.001 & 5.00 & 0.23 & 0.16 & 0.10 & 0.06 & 0.23 & 0.16 & 0.09 & 0.06 & 0.46 & 0.29 & 0.16 & 0.10 & 0.18 & 0.12 & 0.08 & 0.05 \\ 
   & 0.001 & -0.001 & 10.00 & 0.32 & 0.23 & 0.14 & 0.09 & 0.32 & 0.22 & 0.13 & 0.09 & 0.46 & 0.30 & 0.17 & 0.11 & 0.25 & 0.19 & 0.13 & 0.09 \\ 
   & 0.001 & -0.001 & 20.00 & 0.41 & 0.28 & 0.17 & 0.12 & 0.41 & 0.28 & 0.17 & 0.11 & 0.47 & 0.30 & 0.18 & 0.12 & 0.39 & 0.31 & 0.23 & 0.17 \\ 
  \hline
   & 0.0001 & 0.0001 & 1.00 & 0.71 & 0.43 & 0.24 & 0.16 & 0.72 & 0.44 & 0.24 & 0.16 & 2.35 & 1.55 & 0.78 & 0.45 & 0.70 & 0.42 & 0.23 & 0.16 \\ 
   & 0.0001 & 0.0001 & 5.00 & 0.93 & 0.62 & 0.36 & 0.25 & 0.94 & 0.60 & 0.35 & 0.24 & 2.35 & 1.56 & 0.81 & 0.47 & 0.79 & 0.52 & 0.30 & 0.21 \\ 
   & 0.0001 & 0.0001 & 10.00 & 1.24 & 0.89 & 0.55 & 0.38 & 1.22 & 0.85 & 0.53 & 0.38 & 2.38 & 1.61 & 0.86 & 0.52 & 0.97 & 0.68 & 0.43 & 0.30 \\ 
   & 0.0001 & 0.0001 & 20.00 & 1.75 & 1.31 & 0.86 & 0.61 & 1.70 & 1.27 & 0.85 & 0.61 & 2.45 & 1.70 & 1.01 & 0.68 & 1.31 & 0.99 & 0.69 & 0.50 \\ 
   & 0.0001 & -0.0001 & 1.00 & 0.82 & 0.50 & 0.28 & 0.19 & 0.83 & 0.51 & 0.28 & 0.19 & 3.11 & 2.05 & 1.06 & 0.60 & 0.80 & 0.49 & 0.27 & 0.18 \\ 
   & 0.0001 & -0.0001 & 5.00 & 1.19 & 0.80 & 0.48 & 0.33 & 1.17 & 0.76 & 0.46 & 0.32 & 3.13 & 2.07 & 1.09 & 0.64 & 0.94 & 0.63 & 0.38 & 0.27 \\ 
   & 0.0001 & -0.0001 & 10.00 & 1.67 & 1.21 & 0.76 & 0.53 & 1.64 & 1.15 & 0.74 & 0.52 & 3.16 & 2.11 & 1.17 & 0.72 & 1.22 & 0.89 & 0.58 & 0.42 \\ 
   & 0.0001 & -0.0001 & 20.00 & 2.36 & 1.79 & 1.19 & 0.84 & 2.34 & 1.74 & 1.18 & 0.84 & 3.23 & 2.22 & 1.36 & 0.93 & 1.77 & 1.38 & 0.99 & 0.72 \\ 
  \hline
   & 0.00001 & 0.00001 & 1.00 & 7.55 & 4.65 & 2.55 & 1.72 & 7.70 & 4.70 & 2.58 & 1.74 & 27.52 & 18.05 & 9.21 & 5.22 & 7.43 & 4.53 & 2.47 & 1.67 \\ 
   & 0.00001 & 0.00001 & 5.00 & 10.44 & 7.00 & 4.16 & 2.86 & 10.46 & 6.79 & 3.97 & 2.78 & 27.49 & 18.24 & 9.47 & 5.52 & 8.52 & 5.67 & 3.38 & 2.34 \\ 
   & 0.00001 & 0.00001 & 10.00 & 14.34 & 10.38 & 6.48 & 4.51 & 14.09 & 9.90 & 6.28 & 4.45 & 27.85 & 18.73 & 10.20 & 6.22 & 10.84 & 7.70 & 4.97 & 3.56 \\ 
   & 0.00001 & 0.00001 & 20.00 & 20.34 & 15.48 & 10.23 & 7.23 & 20.11 & 15.02 & 10.10 & 7.27 & 28.40 & 19.82 & 11.97 & 8.08 & 15.27 & 11.69 & 8.31 & 6.05 \\ 
   & 0.00001 & -0.00001 & 1.00 & 7.65 & 4.71 & 2.60 & 1.75 & 7.83 & 4.78 & 2.61 & 1.75 & 28.07 & 18.61 & 9.54 & 5.35 & 7.48 & 4.60 & 2.52 & 1.70 \\ 
   & 0.00001 & -0.00001 & 5.00 & 10.72 & 7.20 & 4.28 & 2.93 & 10.67 & 6.95 & 4.06 & 2.85 & 28.30 & 18.73 & 9.79 & 5.69 & 8.70 & 5.77 & 3.46 & 2.41 \\ 
   & 0.00001 & -0.00001 & 10.00 & 14.86 & 10.66 & 6.70 & 4.67 & 14.52 & 10.19 & 6.47 & 4.59 & 28.47 & 19.16 & 10.53 & 6.45 & 11.04 & 7.93 & 5.15 & 3.67 \\ 
   & 0.00001 & -0.00001 & 20.00 & 20.92 & 15.82 & 10.53 & 7.47 & 20.53 & 15.60 & 10.40 & 7.52 & 29.24 & 20.27 & 12.31 & 8.26 & 15.63 & 12.06 & 8.62 & 6.27 \\ 
   \hline
\end{tabular}}}
\end{table}

\FloatBarrier\clearpage
\bibliographystyle{rss}
% \renewcommand{\bibname}{References} 
% \section*{References}
\bibliography{example}
\end{document}